\documentclass[%
 reprint,
 onecolumn,
 12pt,
 amsmath,amssymb,
 aps, 
 nofootinbib]{revtex4-2}

\usepackage{graphicx}
\usepackage{dcolumn}
\usepackage{bm}
\usepackage{algorithm}
\usepackage{xcolor}
\setcitestyle{authoryear,round}
\usepackage{subcaption}

\usepackage{algpseudocode} 

\usepackage{hyperref}

\newif\ifrevhighlight
\revhighlightfalse
\ifrevhighlight
  \newcommand{\rev}[1]{{\color{blue}#1}}
\else
  \newcommand{\rev}[1]{#1}
\fi

\begin{document}

\title{Empathy Modeling in Active Inference Agents for Perspective-Taking and Alignment}

\author{Mahault Albarracin}
 \affiliation{Laboratoire d'ANalyse Cognitive de l'Information, Universite du Quebec a Montreal, Quebec, Canada\\
 Centre of Excellence for AI and Robotics, Sheffield Hallam University, Sheffield, UK
 }

\author{Hongju Pae}
\email{hjpae@activeinference.institute}
\affiliation{Active Inference Institute, California, USA}

\author{Philip Wilson}
\email{philipollaswilson@protonmail.com}
\affiliation{Independent researcher}

\author{Anna Mikeda}
\affiliation{Glass Umbrella, USA}

\author{Alejandro Jimenez Rodriguez}%
 \email{a.jimenez-rodriguez@shu.ac.uk}
\affiliation{%
 Centre of Excellence for AI and Robotics, School of Computing and Digital Technologies\\
 Sheffield Hallam University, Sheffield, UK
}%

\author{Sanjeev V Namjoshi}
\email{sanjeev.namjoshi@gmail.com}
\affiliation{Independent researcher~}

\author{Harshil Shah}
\affiliation {Mission San Jose High School, Fremont, California, USA}

\date{\today}

\begin{abstract}
\rev{Artificial agents that model other agents must predict their behavior and determine whether their outcomes matter within action selection. We introduce an active inference framework that separates these components by combining a history-conditioned Theory of Mind model with an explicit other-regarding valuation parameter, $\lambda$.}
\rev{We instantiate the framework in the Iterated Prisoner's Dilemma. The joint empathy configuration $(\lambda_i,\lambda_j)$ reorganizes the long-run cooperation landscape: sufficiently strong and symmetric other-regarding valuation supports sustained mutual cooperation, whereas strong asymmetry exposes the more empathic agent to systematic exploitation. Along the symmetric diagonal, cooperation exhibits a sharp but continuous finite-precision crossover. Fixed-partner sweeps reveal that the apparent cooperation boundary is path-dependent and that temporal variability is elevated where those paths cross it.}
\rev{Online Bayesian inference over opponent parameters modestly facilitates cooperation near the behavioral boundary but does not substitute for other-regarding valuation. Direct model comparison likewise shows that opponent-sensitive prediction at $\lambda=0$ does not generate cooperation. Planning depth has a partner-dependent effect: it slightly reduces cooperation when modeled reciprocity is weak but strongly increases cooperation against a reciprocating partner such as tit-for-tat. These results distinguish prediction, planning, and prosocial valuation as separable but interacting components of social agency. They also reveal a central limitation of unconditional empathic concern: the same valuation that stabilizes mutual cooperation creates predictable vulnerability when concern is not reciprocated.}
\end{abstract}

\maketitle

\section{\label{sec:level1} Introduction}

\par Artificial agents that can understand and share perspectives are essential for achieving alignment with human values in complex social interactions \citep{dautenhahn1998art}. Traditional approaches to artificial empathy often rely on surface-level pattern recognition and scripted emotional responses \rev{without explicitly representing another agent's latent preferences or social valuation \citep{howcroft2025empathy}. This creates a computational gap between producing socially appropriate responses and inferring the latent variables that organize another agent's behaviour.} To bridge this gap, we propose an active inference framework in which agents treat others’ preferences and social valuation as latent variables to be inferred. Our agents internally model and update beliefs about others' mental states, including their degree of prosocial concern, and \rev{when learning is enabled,} incorporate these beliefs into action selection through an explicit tradeoff between pragmatic and epistemic value, \rev{supporting socially informed behaviour under uncertainty.}

\par We distinguish the components of empathy this framework engages. Cognitive science identifies at least three dissociable facets: cognitive empathy (inferring another's mental states, closely related to theory of mind), affective empathy (resonating with another's emotional states), and a motivational component, empathic concern and the desire to promote another's well-being \citep{weisz2021strategic, decety2004functional, lamm2007neural}. These components are neurally and functionally dissociable \citep{shamay2009two, arioli2021overlapping}. Prior computational work on theory of mind in active inference has engaged the cognitive dimension, primarily predicting what another agent will do. Our framework goes further: we introduce an \textit{empathy parameter} $\lambda$, \rev{an other-regarding valuation parameter, used here to operationalize the empathic-concern component of empathy.} This parameter governs how much weight an agent places on the other's expected free energy during planning. $\lambda$ makes the other's welfare salient within the agent's own decision-making, operationalizing empathic concern within the variational framework, akin to some game-theoretic approaches \citep{orbell1993social, rabin1993incorporating, hwang2018game}. In this sense, our model \rev{captures the motivational component of empathic concern without modelling affective sharing or emotional contagion themselves, and provides a structural placeholder for the motivational dimension. In the present implementation, however, the degree of empathic concern is set exogenously rather than arising from the agent's own need dynamics.} \rev{Throughout the paper, ``empathy'' should therefore be read in this restricted sense: the parameter $\lambda$ implements empathic concern, an other-regarding valuation, while the structurally matched self--other models supply the cognitive substrate of perspective-taking. We return to the components we do not model, such as affective sharing and contagion, in the Discussion.}

\par Our approach draws inspiration from human cognition and neuroscience. In humans, the mirror neuron system supports simulation of others. Observing another's action activates one's own motor and emotional representations \citep{oberman2007simulating}. Similarly, our agent models others using a generative architecture structurally matched to its own, while treating agent-specific parameters as latent variables to be inferred online. Rather than hard-coding opponent characteristics, the agent maintains a posterior distribution over behavioral and empathic parameters, effectively reusing its own cognitive machinery to “step into their shoes” while updating its beliefs through experience. This design aligns with simulation-theoretic accounts of social cognition \citep{goldman2006simulating}, and with theories of second-person neuroscience that emphasize modeling others within one's own cognitive framework \citep{redcay2019using, lehmann2024active}. Prior work has begun integrating theory-of-mind (ToM) into active inference. For example, \citet{demekas2023analytical} model two agents in an Iterated Prisoner’s Dilemma (IPD) as a coupled active inference system, revealing how learning rates and reward structures affect the emergence of cooperative or defection strategies. More recently, \citet{pitliya2025theory} and \citet{ccatal2024belief} demonstrated that active inference agents with explicit ToM achieve improved cooperation without requiring explicit communication, though in some cases through belief sharing.

\par \citet{matsumura2024active} introduce an empathic extension of active inference in which an agent reuses a structurally matched generative model to simulate another agent’s perspective, consistent with simulation-theoretic accounts of social cognition. In their formulation, the agent incorporates estimates of the other’s internal state into policy evaluation and can select actions that reduce the other agent’s expected free energy, promoting socially appropriate behavior in embodied navigation tasks. Their implementation is grounded in domain-specific dynamics (e.g., social-force models for multi-robot navigation) and focuses on improving coordination and safety margins in situated environments. While this work demonstrates that active inference can support empathic behavior in embodied settings, it does not examine repeated strategic interaction under simultaneous decision-making, nor does it analyze \rev{long-run interaction regimes}, exploitation asymmetry, or regime-boundary dynamics. In contrast, our framework embeds empathic valuation within a formally specified game-theoretic setting, introduces latent inference over opponent valuation parameters, and characterizes how \rev{other-regarding valuation reshapes behavioural stability, threshold behaviour, and vulnerability to exploitation} in iterated dilemmas.

\par More broadly, prior active inference approaches to social interaction typically instantiate separate self- and other-models but do not treat opponent valuation itself as a hidden variable subject to epistemic inference, nor do they analyse how such latent social parameters alter \rev{interaction-regime} selection and dynamical stability. These approaches also remain within the cognitive empathy domain; they model what another agent will do, but not \rev{whether that agent's outcomes carry value within the modelling agent's own objective}.

\par Our contribution is a unified algorithmic framework \rev{combining Theory of Mind with other-regarding valuation} in active inference, in which each agent maintains a generative model structurally matched to its own architecture when modeling others. Rather than hand-coding discrete opponent types, agent $i$ represents agent $j$'s behavioral and valuational parameters as latent variables and performs Bayesian inference over them online. This preserves a shared generative structure, common state space, transition dynamics, and observation mappings, while allowing agent-specific parameters governing cooperation bias, reciprocity, precision, and empathic valuation to be inferred from interaction history.

\par Perspective-taking is realized as a continuous empathy-weighted blend of self- and other-oriented expected free energy:
\begin{equation}
\mathbf{G}_{\text{social}}(a_i)
=
(1 - \lambda)\,
\mathbf{G}_{\text{self}}(a_i \mid q(a_j \mid h_t))
+
\lambda\,
\rev{\mathbf{G}_{\text{other}}(a_i \mid q(a_j \mid h_t))}
\end{equation}
where $\lambda \in [0,1]$ controls the degree of empathic concern \rev{and $q(a_j \mid h_t)$ is the predicted distribution over the opponent's next action. In the learning-enabled variant, the opponent's empathic weighting is itself treated as a hidden variable, allowing uncertainty about $\lambda_j$ to contribute an epistemic term to action evaluation.} \rev{This architecture separates two questions that are often run together: whether an agent models what its partner will do, and whether the partner's outcomes carry weight in its own objective. We test this distinction directly in Section~\ref{sec:model_comparison}: enabling learned opponent inversion while holding $\lambda=0$ does not materially increase cooperation relative to the corresponding static-ToM, self-regarding agent, whereas the same inversion machinery combined with sufficiently strong other-regarding valuation sustains cooperation. Thus, cooperation in the present experiments does not follow from learned opponent modelling alone; instead, it requires the partner's outcomes to enter the focal agent's valuation.} We integrate this mechanism into the active inference perception-action loop under simultaneous decision-making. \rev{In the learning-enabled condition, action selection combines pragmatic and epistemic components of expected free energy. Across the examined Iterated Prisoner's Dilemma conditions, the resulting architecture supports sustained mutual cooperation at sufficiently strong other-regarding valuation, whereas self-regarding variants predominantly defect.}

\section{Methodology}\label{sec:methodology}

\subsection{Generative Model for an Empathic Agent}
We formalize each agent's belief system as a \emph{partially observable Markov decision process} (POMDP) generative model, extended to support perspective-taking as demonstrated in \citet{pitliya2025theory}. The agent maintains a probability distribution over latent hidden states, sensory observations, and actions/policies over time. Let $s_\tau$ denote the hidden state at time $\tau$, $o_\tau$ the observation at time $\tau$, and $\pi$ a policy (a sequence of future control states $u_\tau$). Under standard Markovian assumptions, the joint generative model factorizes as
\begin{equation}
P(\tilde{o}, \tilde{s}, \pi)
=
P(s_0)\,P(\pi)
\prod_{\tau=1}^{T} P\!\left(s_\tau \mid s_{\tau-1}, u_\tau\right)
\prod_{\tau=0}^{T} P\!\left(o_\tau \mid s_\tau\right)
\label{eq:generative_model_pomdp}
\end{equation}

where $\tilde{s} = (s_0, \dots ,s_T)$ and $\tilde{o} = (o_0, \dots ,o_T)$ denote state and observation trajectories, respectively.

Model parameters include the state-transition and observation likelihood mappings, commonly represented in active inference by tensors (or matrices) $B$ and $A$, respectively, along with preferences over outcomes (prior over observations) $C$ and initial state priors $D$. All random variables are discrete in our implementation. We further distinguish \emph{first-order} parameters (e.g., the entries of $A, B, C, D$) from \emph{second-order} parameters that modulate learning rates and/or precision (inverse temperature) of the corresponding distributions.

Posterior inference is performed via variational Bayes. The agent maintains a recognition density $Q(s_\tau)$ that approximates the true posterior and updates this density by minimizing \emph{variational free energy} (VFE). Intuitively, this corresponds to continually updating beliefs $Q(s_\tau)$ so as to better explain incoming observations under the agent's generative model.

\subsection{Hidden state factorization and observation modalities}

In our model, each agent maintains a generative model over joint interaction outcomes. The hidden state $s_\tau$ is a single multinomial latent variable encoding the joint outcome of the current round:                                                                  
  \begin{equation}
      s_\tau \in \{\text{CC},\; \text{CD},\; \text{DC},\; \text{DD}\}
  \end{equation}
where C and D denote cooperation and defection, and the pair indicates (my action, opponent's action). This state space jointly encodes both the agent's own behaviour and the opponent's, providing the minimal sufficient representation for computing payoffs and planning future actions in a simultaneous-move game.
The observation model is an identity mapping $A = I_4$, such that the agent directly observes the joint outcome at each round. This means the agent has full access to both its own and the opponent's realised actions, but must infer the opponent's \emph{future} behaviour through Theory of Mind (Section~\ref{sec:tom}).

Rather than introducing a binary latent variable to switch between egocentric and allocentric perspectives, our model implements perspective-taking as a \emph{continuous blend} of self- and other-oriented expected free energy. For each candidate action $a_i$, the agent computes a social EFE:
\begin{equation}
\begin{aligned}
\mathbf{G}_{\text{social}}(a_i)
&= (1 - \lambda)\, \mathbf{G}_{\text{self}}(a_i \mid q(a_j \mid h_t)) \\
&\quad + \lambda\, 
\rev{\mathbf{G}_{\text{other}}
(a_i \mid q(a_j \mid h_t))}
\end{aligned}
\label{eq:social_efe}
\end{equation}
where $\lambda \in [0, 1]$ is the empathy parameter and $q(a_j \mid h_t)$ denotes the predicted opponent response conditioned on interaction history $h_t$. The empathy parameter $\lambda$ is a fixed architectural parameter that determines the degree to which the agent weighs the other's welfare in its own planning. At $\lambda = 0$, the agent is purely self-interested; at $\lambda = 1$, it is purely ``altruistic'' \rev{(i.e. action evaluation is based entirely on the opponent's expected payoff)}; intermediate values produce a graded trade-off between egocentric and allocentric evaluation.

The opponent's predicted response $q(a_j \mid h_t)$ is obtained from a depth-1 Theory of Mind module. \rev{In the static case, the opponent is modeled as selecting actions under an estimated distribution over the focal agent's behaviour, and the prediction takes the form:}
\begin{equation}
q(a_j \mid h_t) \propto
\exp\bigl(-\beta_j\, \mathbf{G}_j(a_j \mid h_t)\bigr)
\end{equation}
where $\beta_j$ is the opponent's precision and $\mathbf{G}_j(a_j \mid h_t)$ is the opponent's expected free energy \rev{under its estimated distribution over the focal agent's actions, derived from the observed interaction history}. \rev{When opponent inversion is active, the static prediction is combined with a posterior-predictive distribution obtained through Bayesian updating over latent behavioral parameters, and focal action evaluation additionally includes epistemic value (see Section~\ref{sec:learning}).}

The transition model $B$ is parameterized as a set of action-conditional matrices $B^{(a_i)} \in \mathbb{R}^{4 \times 4}$, one for each of the agent's two actions (cooperate or defect). Because the joint outcome also depends on the opponent's action, which is not under the agent's control, the transition matrices encode stochastic mappings that
reflect uncertainty over the opponent's response, distributing probability mass across the two possible outcomes consistent with each of the agent's actions.

Each agent's preferences are encoded in the vector $C \in \mathbb{R}^4$, which assigns a log-preference (or negative surprise) to each joint outcome. \rev{The PyMDP agents are instantiated with preference vectors corresponding to the Prisoner's Dilemma outcome ordering. In the Theory of Mind-based experiments reported here, however, behavioral action selection is performed by the social-EFE module described below, which computes expected self- and other-payoffs directly from the decision payoff matrix $(R,S,T,P)=(3,0,5,1)$. The $C$ vectors therefore remain part of the PyMDP generative model specification, but do not themselves determine the actions selected through the custom ToM pathway.}

\subsection{Perspective-taking via structurally matched generative models}

Central to our approach is the assumption that an agent models others using a generative architecture structurally matched to its own \citep{ford2012executive}. Let 
$M_{\text{self}}^i = (A_i, B_i, C_i, D_i)$ denote agent $i$'s generative model of itself. 
When modeling agent $j$, the agent does not construct an arbitrary independent model; rather, it instantiates a generative model with identical structural form (state space, observation model, and inference machinery), while treating opponent-specific behavioral and valuational parameters as latent variables to be inferred online.

Formally, we denote the opponent model as
\begin{equation}
    M_{\text{other}}^i =
    \rev{(A_j, B_j, C_j, D_j;\theta_j)}
\end{equation}
where $\theta_j$ comprises latent parameters governing cooperation bias, reciprocity, precision, and empathic weighting. These parameters are updated via Bayesian inference over interaction history.

The resulting other-model shares the same structural form as the self-model: the same state and observation dimensions and the same PyMDP inference machinery. It differs in its parameterization, which is inferred rather than directly observed. This construction is inspired by simulation theory, according to which an agent understands others by reusing its own cognitive architecture under alternative parameter settings \citep{goldman2006simulating, gallese1998mirror}.

In our implementation, the self-model and other-model are instantiated as separate but structurally matched generative models at the beginning of an interaction. The self-model $M_{\text{self}}^i$ is used for state inference, while the other-model $M_{\text{other}}^i$ is passed to a Theory of Mind module that predicts the opponent's likely response based on interaction history (Section~\ref{sec:tom}). \rev{Opponent actions are predicted by the dedicated Theory of Mind and opponent-inversion modules; the structurally matched PyMDP models provide the shared state and outcome representation rather than a recursive simulation of PyMDP policy inference.}

Perspective-taking does not occur via a latent mode-switching variable; instead, egocentric and allocentric evaluations are computed simultaneously and blended through the social expected free energy (social EFE):
\begin{equation}
    \mathbf{G}_{\text{social}}(a_i)
    = (1 - \lambda)\,
    \mathbf{G}_{\text{self}}(a_i \mid q(a_j \mid h_t)) 
    + \lambda\, 
    \rev{\mathbf{G}_{\text{other}} (a_i \mid q(a_j \mid h_t))}
    \label{eq:g_social}
\end{equation}
where $\lambda \in [0,1]$ is the empathy parameter and $q(a_j \mid h_t)$ denotes the predicted opponent response conditioned on interaction history $h_t$.

\rev{Here, both $\mathbf{G}_{\text{self}}$ and $\mathbf{G}_{\text{other}}$ are expected negative payoffs under the predicted opponent-action distribution, evaluated from the focal and opponent perspectives, respectively. In particular,
\begin{equation}
    \mathbf{G}_{\text{other}}
    (a_i \mid q(a_j \mid h_t))
    =
    -\sum_{a_j}
    q(a_j \mid h_t)\,
    \mathrm{payoff}_j(a_i,a_j)
\label{eq:g_other}
\end{equation}}

\rev{The static opponent prediction is defined in Eq.~\eqref{eq:opponent_G}. The predictive quantity $G_j(a_j \mid h_t)$ evaluates the opponent's expected negative payoff for choosing $a_j$ under its estimated distribution over the focal agent's actions. This quantity is distinct from $\mathbf{G}_{\text{other}}(a_i \mid q(a_j \mid h_t))$, which evaluates the opponent's expected negative payoff conditional on the focal agent's candidate action $a_i$. When opponent inversion is active, the static prediction is combined with the posterior-predictive distribution over latent opponent parameters, and epistemic value is included in focal action selection.}

\rev{A natural question is why empathic concern is a single bipolar scalar rather than separate self- and other-concern weights, $G = w_{\text{self}}\mathbf{G}_{\text{self}} + w_{\text{other}}\mathbf{G}_{\text{other}}$. On the positive quadrant the two are equivalent up to precision. Because actions are selected by a softmax with precision $\beta$, $\beta(w_{\text{self}}\mathbf{G}_{\text{self}} + w_{\text{other}}\mathbf{G}_{\text{other}}) = \beta'[(1-\lambda)\mathbf{G}_{\text{self}} + \lambda\mathbf{G}_{\text{other}}]$ with $\lambda = w_{\text{other}}/(w_{\text{self}}+w_{\text{other}})$ and $\beta' = \beta(w_{\text{self}}+w_{\text{other}})$: the direction of the weight vector fixes the empathy parameter and its magnitude simply rescales action precision. We confirmed this in simulation, where rescaling both weights by $c$ reproduces the trajectories of the bipolar model with precision $c\beta$ seed for seed. The two-weight parameterisation also represents regimes outside the positive quadrant that $\lambda \in [0,1]$ cannot reach: negative $w_{\text{other}}$ represents spite (the other's loss is valued), whereas negative $w_{\text{self}}$ represents self-abnegation. We examine these sign-extended regimes in Section~\ref{sec:model_comparison} and Figure~\ref{fig:fig10_weight_space}.}

This design has several important consequences. First, shared structural assumptions ensure that environmental dynamics and observation mappings remain aligned across perspectives. Second, the continuous empathy parameter $\lambda$ provides a smooth interpolation between egocentric and allocentric evaluation. Third, the separation of \emph{what the opponent will do} (Theory of Mind inference) from \emph{how much to care} (empathic weighting) allows each component to be analyzed independently.

\subsection{Active inference and sophisticated planning}
\label{sec:method_sophistication}
\label{sec:tom}

Having specified the generative model for each agent, we now describe the inference and planning procedure that governs agent behaviour, inspired by sophisticated inference as described in \cite{friston2021sophisticated}. Each agent operates in discrete perception–action cycles, alternating between variational state inference, opponent modelling, and policy evaluation. An overview is given in Algorithm~\ref{alg:aif_perspective}.
At each time step $\tau$, the agent receives a joint-outcome observation $o_\tau \in {\text{CC}, \text{CD}, \text{DC}, \text{DD}}$ and updates its posterior beliefs $Q(s_\tau)$ over the four-state hidden variable by minimising variational free energy. This filtering step combines the predictive prior from the previous posterior and the transition model $\mathbf{B}$ with the observation likelihood $\mathbf{A}$, using standard active inference state inference \citep{dacosta2020active}.
In parallel with state inference, the agent maintains a parametric model of its opponent via a particle-based inversion module. Rather than enumerating discrete strategy types, each particle $k$ carries a parameter vector
\[
\theta_k = (\alpha_k, \rho_k, \beta_k, \lambda_{j,k})
\]
where $\alpha$ encodes cooperation bias, $\rho$ reciprocity sensitivity, $\beta$ behavioral precision, and $\lambda_j$ the opponent's empathic weighting. 

The opponent's probability of cooperation is modeled as:
\begin{equation}
P(a_j = C \mid h_t, \theta_k)
=
\sigma\!\bigl(\beta_k(\alpha_k + \rho_k f(h_t) + \text{empathy\_shift}(\lambda_{j,k},p))\bigr)
\end{equation}
where $\sigma$ is the logistic function. \rev{Because actions are simultaneous, the reciprocity feature is aligned to the information available when the observed opponent action was selected. For an opponent action observed at round $t-1$, $f(h_t)$ equals $+1$ if the focal agent cooperated at round $t-2$, $-1$ if it defected, and $0$ when this lagged action is unavailable. The empathy shift is the social-EFE cooperation advantage for an opponent with empathic weighting $\lambda_{j}$ facing an agent whose cooperation probability it estimates as $p$: under the decision payoffs $(R,S,T,P)=(3,0,5,1)$, $\text{empathy\_shift}(\lambda_{j},p) = 5\lambda_{j} - p - 1$. At $\lambda_{j}=0$, the empathy-shift term reduces to $-(p+1)$ and therefore favours defection when the remaining terms are held fixed. This term crosses zero at $\lambda_{j}=(p+1)/5$, which for $p \in [0,1]$ yields the interval $\lambda_j \in [0.2,0.4]$. The resulting action probability also depends on cooperation bias $\alpha_k$, reciprocity $\rho_k$, precision $\beta_k$, and interaction history.}

Particle weights are updated via Bayesian filtering:
\begin{equation}
w_k^{(\tau)} \propto
w_k^{(\tau-1)}
P(a_j^{(\tau)} \mid h_t, \theta_k)
\end{equation}

The posterior-weighted ensemble yields the predictive distribution:
\begin{equation}
q_{\text{learned}}(a_j \mid h_t)
=
\sum_k w_k P(a_j \mid h_t, \theta_k)
\end{equation}
Because the opponent's empathic weighting $\lambda_j$ is treated as a hidden variable, action selection incorporates epistemic value. For candidate action $a_i$, the epistemic component is defined as
\begin{equation}
G_{\text{epistemic}}(a_i)
=
- \mathbb{E}\!\left[
\mathrm{KL}\bigl(
\rev{p(\lambda_j \mid a_j^{(\tau+1)}, h_t, a_i)}
\;\|\;
\rev{p(\lambda_j \mid h_t)}
\bigr)
\right]
\end{equation}
which quantifies the expected information gain about the opponent's \rev{empathic weighting $\lambda_j$. In the implementation, the marginal posterior over $\lambda_j$ is approximated by binning the weighted particles, and epistemic value is computed as the expected reduction in this marginal entropy.}

Because the particle filter may be unreliable early in an interaction (when few observations have been collected), the agent gates between the learned posterior-predictive distribution and a static Theory of Mind (ToM) prior. In the static ToM, the opponent is modeled as selecting actions based only on interaction history $h_t$ (rather than on the agent's current-round action), consistent with simultaneous decision-making. The static ToM distribution is given by
\begin{equation}
    q_{\text{static}}(a_j \mid h_t)
    \propto
    \exp\!\left(-\beta_j\, G_j(a_j \mid h_t)\right)
    \label{eq:opponent_G}
\end{equation}
where $\beta_j$ is the assumed opponent precision and $G_j(a_j \mid h_t)$ is the opponent's expected negative payoff for choosing $a_j$, computed under the opponent's inferred policy over the agent's actions (estimated from history). Concretely, we form an empirical belief $\pi_i(a_i \mid h_t)$ over the agent's own action (e.g., based on past cooperation rate), and evaluate
\begin{equation}
    G_j(a_j \mid h_t)
    =
    -\sum_{a_i \in \{C,D\}} \pi_i(a_i \mid h_t)\,\text{payoff}_j(a_i,a_j)
\end{equation}

The gated prediction smoothly interpolates between learned and static predictions based on the inversion reliability $r \in [0,1]$, computed from \rev{agreement among the particles' predicted opponent-action probabilities}:
\begin{equation}
    q_{\text{gated}}(a_j \mid h_t)
    =
    r \cdot q_{\text{learned}}(a_j \mid h_t)
    +
    (1-r)\cdot q_{\text{static}}(a_j \mid h_t)
    \label{eq:gated_prediction}
\end{equation}
\rev{When particle predictions disagree, reliability is low and the agent falls back toward the static ToM prior; when the particle ensemble converges on a common prediction, reliability is high and the learned model dominates.} \rev{The static and learning-enabled variants compared throughout the paper are therefore the same agent with one module toggled. Both share the generative model, state inference, social EFE, and action selection described above; they differ only in whether the particle-filter inversion is active. With inversion disabled the opponent prediction reduces to the static prior $q_{\text{static}}$, and with it enabled the gated mixture above replaces it, with all remaining parameters held identical. Epistemic value is included only in the inversion-enabled variant.}

Following state inference and opponent modelling, the agent evaluates candidate policies. We implement two planning regimes. In the \emph{myopic} regime ($H = 1$), the agent evaluates a single-step total expected free energy for each candidate action $a_i \in \{C,D\}$ as
\begin{equation}
\begin{aligned}
    G_{\text{total}}(a_i)
    &=
    (1-\lambda)\,G_{\text{self}}(a_i \mid q(a_j \mid h_t))
    +
    \lambda\,\rev{G_{\text{other}} (a_i \mid q(a_j \mid h_t))} \\
    &\quad +\, G_{\text{epistemic}}(a_i)
\end{aligned}
\label{eq:gtotal_myopic}
\end{equation}
where $G_{\text{self}}(a_i \mid q(a_j \mid h_t)) = -\mathbb{E}_{q(a_j \mid h_t)}[\text{payoff}_i(a_i,a_j)]$ is the agent's expected negative payoff under the predicted opponent response, \rev{and $G_{\text{other}}(a_i \mid q(a_j \mid h_t))$ is the opponent's expected negative payoff conditional on candidate focal action $a_i$, averaged under the predicted opponent-action distribution. When opponent inversion is active, $G_{\text{epistemic}}(a_i)$ is the one-step expected information gain about $\lambda_j$ defined above. When inversion is disabled, this term is set to zero.}

In the \emph{sophisticated} regime ($H > 1$), the agent enumerates all $2^H$ candidate policies $\pi = (a_i^{(0)}, a_i^{(1)}, \ldots, a_i^{(H-1)})$ and evaluates each by forward rollout. At each rollout step $t$, the opponent's response is predicted and the total expected free energy is accumulated:
\begin{equation}
    G(\pi)
    =
    \sum_{t=0}^{H-1}
    \Big[
    (1-\lambda)\,G_{\text{self}}^{(t)}
    +
    \lambda\,G_{\text{other}}^{(t)}
    +
    G_{\text{epistemic}}^{(t)}
    \Big]
    \label{eq:g_policy}
\end{equation}
\rev{The horizon sum is not normalised by $H$. Because the policy softmax applies $\beta$ to $G(\pi)$, dividing by $H$ rescales the effective action precision to $\beta/H$. If the rollout returns the same opponent prediction for every candidate policy, the resulting policy distribution is then exactly a myopic choice at precision $\beta/H$, so the horizon acts as a temperature parameter rather than as lookahead. We note the equivalence explicitly because this normalisation is common in implementations of sophisticated inference.}
At rollout step $t = 0$, the gated prediction $q_{\text{gated}}(a_j \mid h_t)$ is used; at future steps $t>0$, the opponent prediction is conditioned on the simulated history induced by the partial rollout (i.e., past actions in the simulated future). The agent then forms a posterior over policies:
\begin{equation}
    Q(\pi) \propto \exp\!\left(-\beta\,G(\pi)\right)
\end{equation}
and marginalises to the first action:
\begin{equation}
    P(a_i^{(0)} = a) = \sum_{\pi:\,\pi_0=a} Q(\pi)
\end{equation}
An action is sampled from this distribution and executed. The perception--action cycle repeats at $\tau + 1$.

\begin{algorithm}[H]
\caption{Active inference with empathic planning}
\label{alg:aif_perspective}
\begin{algorithmic}[1]
\State Initialise beliefs $Q(s_0)$ with prior $\mathbf{D}$; initialise particle filter with uniform weights over $\theta_j$
\For{$\tau = 0$ to $T-1$}
    \State Observe joint outcome $o_\tau$
    \State Extract opponent action $a_j^{(\tau)}$ from $o_\tau$
    \State Update particle weights: $w_k \leftarrow w_k \cdot P(a_j^{(\tau)} \mid h_t,\theta_k)$; normalise
    \State Update beliefs $Q(s_\tau)$ via variational state inference
    \State \rev{Compute reliability $r$ from agreement among particle predictions}
    \State Form predictions: $q_{\text{learned}}(a_j \mid h_t)=\sum_k w_k P(a_j \mid h_t,\theta_k)$
    \State Compute static ToM prior $q_{\text{static}}(a_j \mid h_t)$ via Eq.~\eqref{eq:opponent_G}
    \State Gate predictions: $q_{\text{gated}}(a_j \mid h_t) \leftarrow r\,q_{\text{learned}} + (1-r)\,q_{\text{static}}$
    \For{each candidate policy $\pi = (a_i^{(0)}, \ldots, a_i^{(H-1)})$}
        \For{$t = 0$ to $H-1$}
            \State Predict opponent response $q(a_j \mid h_t^{(t)}) \leftarrow$ \textbf{if} $t=0$: $q_{\text{gated}}(a_j \mid h_t)$; \textbf{else}: $q_{\text{static}}(a_j \mid h_t^{(t)})$
            \State Compute $G_{\text{self}}^{(t)}$ and $G_{\text{other}}^{(t)}$ under $q(a_j \mid h_t^{(t)})$
            \State Compute epistemic term $G_{\text{epistemic}}^{(t)}$ (one-step expected information gain)
            \State Accumulate $G^{(t)} \leftarrow (1-\lambda)G_{\text{self}}^{(t)} + \lambda G_{\text{other}}^{(t)} + G_{\text{epistemic}}^{(t)}$
        \EndFor
        \State $G(\pi) \leftarrow \sum_t G^{(t)}$
    \EndFor
    \State $Q(\pi) \propto \exp(-\beta\,G(\pi))$
    \State Marginalise: $P(a_i^{(0)}=a) = \sum_{\pi:\,\pi_0=a} Q(\pi)$
    \State Sample and execute $a_i^{(\tau)} \sim P(a_i^{(0)})$
\EndFor
\end{algorithmic}
\end{algorithm}

\section{Results}\label{sec:results}                                                            
  \subsection{Iterated Prisoner's Dilemma setup and global cooperation landscape}

    \par We first characterize the global cooperation landscape induced by empathic weighting in the Iterated Prisoner's Dilemma (IPD). \rev{Across dyads, the pair $(\lambda_i,\lambda_j)$ defines a joint control space that reshapes long-run interaction regimes. Along the symmetric diagonal, where $\lambda_i=\lambda_j=\lambda$, increasing empathic weighting induces a sharp shift from mutual defection to sustained cooperation.}
    
    \par We evaluate the proposed model in the classic two-player IPD, a canonical paradigm for studying the emergence of cooperation under strategic tension \citep{rapoport1965ipd}. In each round, both agents simultaneously select either cooperate ($C$) or defect ($D$). Payoffs follow the standard ordering $T > R > P > S$ (concretely $T=5$, $R=3$, $P=1$, $S=0$), where $R$ is the mutual cooperation reward $(C,C)$, $T$ is the temptation payoff for unilateral defection $(D,C)$, $P$ is the mutual defection punishment $(D,D)$, and $S$ is the sucker's payoff $(C,D)$. This payoff structure captures the core dilemma that unilateral defection is individually advantageous ($T>R$), yet mutual cooperation yields higher collective welfare than mutual defection ($R>P$). Unless otherwise stated, simulations in this subsection use myopic planning ($H=1$), \rev{with opponent inversion enabled and symmetric precision parameters. Each interaction consists of 100 repeated rounds, with results averaged over 20 random seeds.}

    \par Each agent is parameterized by an empathy level $\lambda \in [0,1]$, which enters the social EFE as: 
    \begin{equation}
    G_{\text{social}}(a_i)
    =
    (1-\lambda)\,G_{\text{self}}(a_i \mid q(a_j \mid h_t))
    +
    \lambda\,\rev{G_{\text{other}}(a_i \mid q(a_j \mid h_t))}
\end{equation}
    Here, setting $\lambda=0$ yields a purely self-oriented agent, whereas $\lambda=1$ yields a fully other-oriented agent. Intermediate values implement a graded tradeoff between self-interest and prosocial concern. 

   \begin{figure}[tbp]
      \centering
      \captionsetup{font=footnotesize}
      \includegraphics[width=1.0\linewidth]{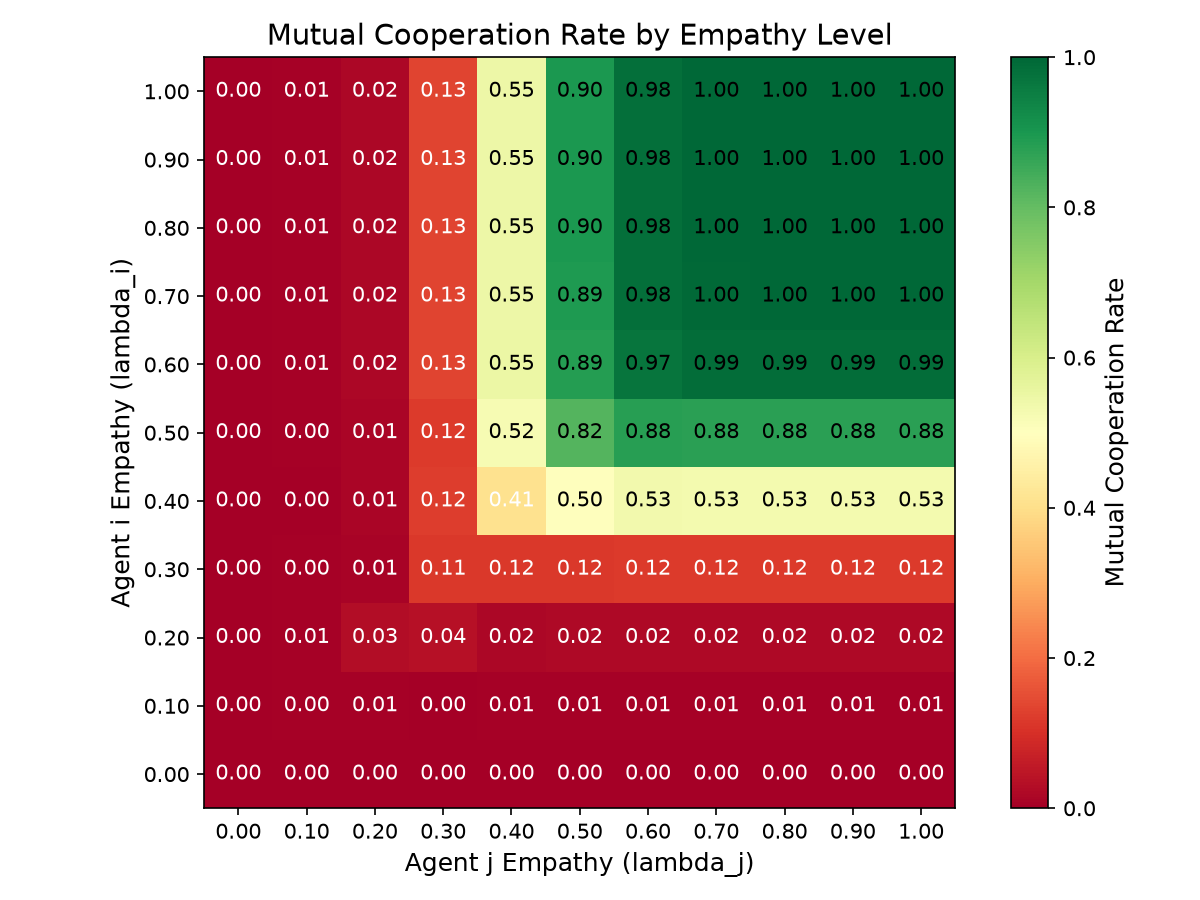}
      \caption{Mutual cooperation landscape across dyadic empathy. 
      Each cell shows the mean fraction of rounds ending in mutual cooperation $(C,C)$ for the corresponding empathy pair $(\lambda_i,\lambda_j)$, \rev{averaged over 20 seeds and 100 rounds per interaction}. Axes indicate the empathy parameters of the two agents $\lambda_i$ and $\lambda_j$.} 
      \label{fig:fig1_agent_empathy}
  \end{figure}

    \par Figure~\ref{fig:fig1_agent_empathy} shows the fraction of rounds resulting in mutual cooperation $(C,C)$ across the $(\lambda_i,\lambda_j)$ grid. A clear regime structure emerges: \rev{at low values of other-regarding valuation, mutual cooperation remains rare and interactions remain concentrated in mutual defection. Along the symmetric diagonal, mutual cooperation crosses $P(CC)=0.5$ near $\lambda=0.40$, exceeds $0.8$ by $\lambda=0.50$, and approaches near-perfect cooperation at higher values.}
    
    \par Moderate increases in empathy around the threshold produce disproportionately large changes in cooperation frequency. \rev{The resulting landscape is consistent with a sharp finite-precision crossover in which increasing empathic weighting $\lambda$ shifts the predominant long-run interaction regime from $(D,D)$ to $(C,C)$.} 
    
    \par Given these configurations, we next examine (1) the role of empathy symmetry and exploitation under asymmetric dyads, (2) temporal recovery dynamics and implicit communication, \rev{(3) boundary-layer variability along fixed-partner paths through the joint empathy space, (4) an explicit characterization of the symmetric-diagonal transition point $\lambda^*$, (5) the effects of online opponent inference on cooperation, (6) the dissociation between opponent prediction and other-regarding valuation, and (7) the consequences of multi-step planning under different partner models.}

    \subsection{Emergent exploitation dynamics}

\par While \rev{sufficiently high empathy in symmetric dyads} supports stable cooperation, \rev{sufficiently strong asymmetries in empathic weighting} give rise to systematic \textit{exploitation}. We therefore examine payoff outcomes under empathy imbalance. 

    \begin{figure}[H] 
    \centering 
    \captionsetup{font=footnotesize}
    \includegraphics[width=1.0\linewidth]{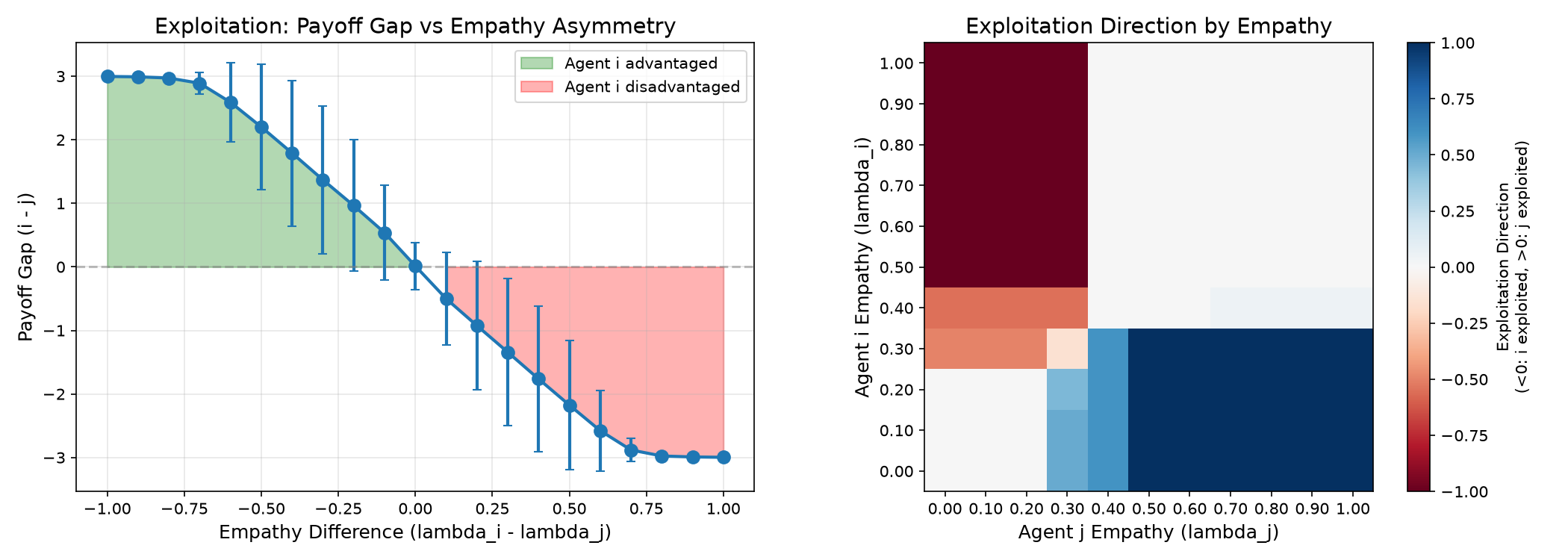} 
    \caption{\rev{Strong} empathy asymmetry induces systematic exploitation. Mean payoff gap as a function of empathy difference $\lambda_i-\lambda_j$, where positive values indicate that agent $i$ obtains higher average payoff. Error bars denote $\pm1$ standard deviation across runs aggregated within each rounded empathy-difference bin. \rev{Results are computed over 100-round interactions with opponent inversion enabled.}}
    \label{fig:fig2_exploitation_payoff} 
    \end{figure}

\par Figure~\ref{fig:fig2_exploitation_payoff} summarizes the resulting dynamics. \rev{Across asymmetric pairings, the payoff advantage is directed toward the less empathic agent and becomes increasingly pronounced as the disparity grows.} When $\lambda_i \ll \lambda_j$, agent $i$ assigns negligible weight to the other's welfare and defects more frequently, while agent $j$, heavily weighting $G_{\text{other}}$, continues to cooperate despite being exploited. This produces a positive payoff gap in favor of the less empathic agent. The direction reverses when $\lambda_i \gg \lambda_j$, \rev{producing the mirror-image pattern of exploitation}. 

\par The magnitude of the payoff gap increases approximately monotonically with empathy disparity over the tested range. This exploitation effect arises as a structural consequence of the social EFE formulation: increasing $\lambda$ progressively reshapes the agent's objective toward other-regarding valuation, thereby altering the stability of unilateral defection. In the limiting case $(\lambda_i,\lambda_j)=(0,1)$, the interaction corresponds to the ``sucker'' outcome of the Prisoner's Dilemma. \rev{Consistent with this limit, in the strongly asymmetric pairing of $(\lambda_i,\lambda_j)=(0.6,0)$, the more empathic agent cooperates while its self-regarding partner defects, yielding an exploitability of $1.00$ and a mean per-round payoff gap of $-5.00$ from the more empathic agent's perspective.}

\par These dynamics arise under simultaneous decision-making with history-conditioned opponent prediction; \rev{exploitation does not depend on observing or conditioning on the partner's current-round action but follows directly from asymmetric valuation.} \rev{These results show that other-regarding valuation stabilizes cooperation only when it is sufficiently mutual. In the absence of such mutuality, empathic concern creates predictable vulnerability. Opponent inversion can track partner asymmetries, but with the focal empathy parameter $\lambda$ fixed, improved prediction does not itself provide a mechanism for withdrawing cooperation under persistent exploitation. This structural tension motivates the learning and model comparison analyses in Sections~\ref{sec:learning} and~\ref{sec:model_comparison}.}

    \subsection{Implicit communication and recovery dynamics}
    
    \par Beyond \rev{aggregate cooperation frequencies}, the temporal dynamics of interaction reveal \rev{action-mediated coordination under high empathy, which we interpret as a minimal form of \textit{implicit communication}}. In our model, agents influence one another solely through their action choices, and their behavior \rev{becomes increasingly aligned over repeated interaction}.
    
    \par
\begin{figure}[tbp]
    \centering
    \captionsetup{font=footnotesize}
    \includegraphics[width=1.0\textwidth]{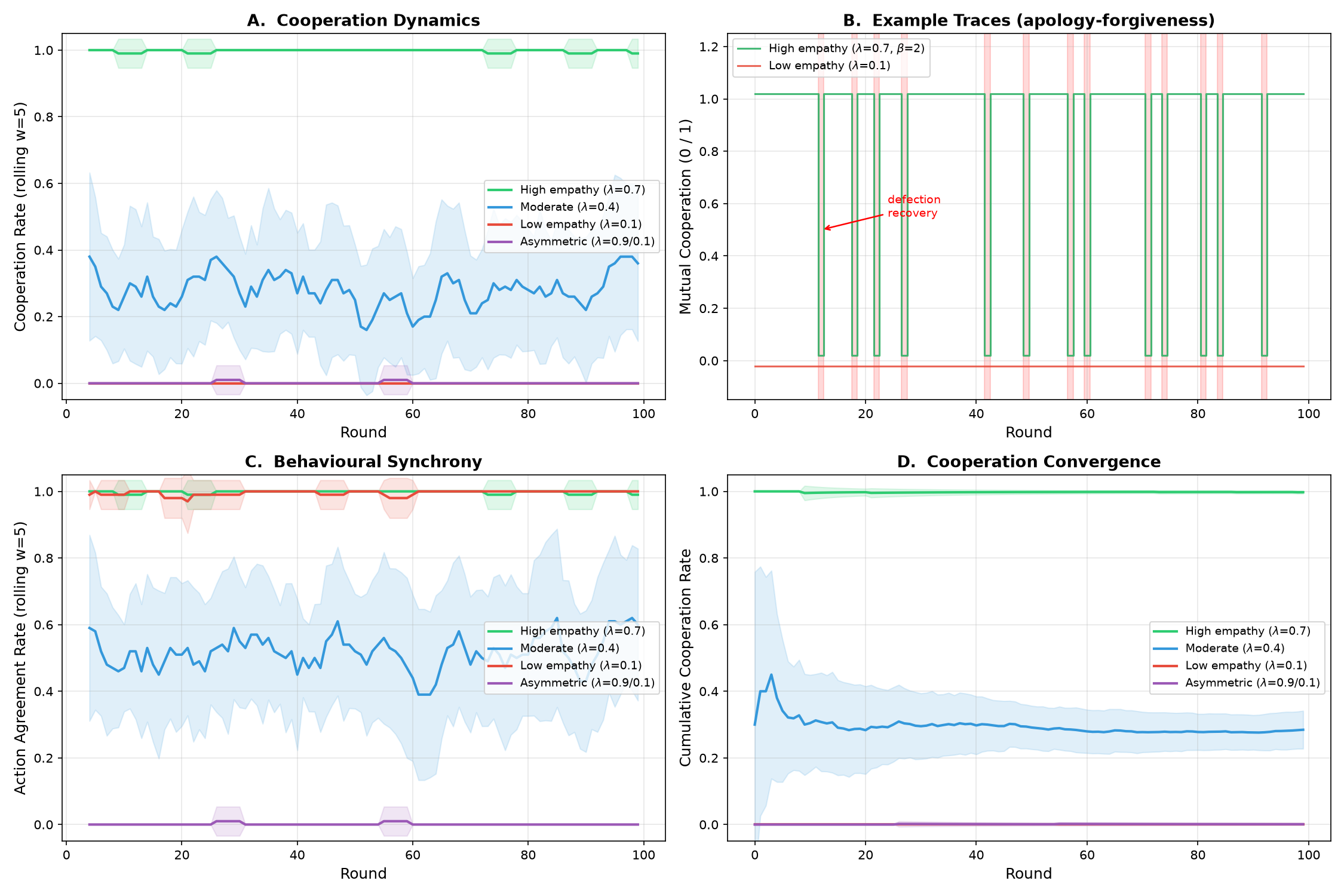}
    \caption{\textbf{Temporal cooperation dynamics across empathy regimes.} All panels summarize 100-round interactions \rev{under fixed empathic weighting and with opponent inversion disabled, thereby isolating history-conditioned static-ToM dynamics. Panels A, C, and D show means across 20 seeds under a rolling statistics window of $w=5$, with shaded bands denoting $\pm1$ standard deviation.} 
    \textbf{A:} Rolling mutual cooperation rate. High empathy ($\lambda=0.7$) rapidly converges to and stabilizes near-perfect cooperation, \rev{whereas moderate empathy ($\lambda=0.4$), which lies near the transition region, exhibits intermediate and more variable cooperation}. Low empathy ($\lambda=0.1$) and strongly asymmetric empathy ($\lambda=0.9/0.1$) remain near persistent defection. 
    \textbf{B:} Example single-seed trajectories highlighting recovery (apology-forgiveness). \rev{The high-empathy trace uses a lower-precision condition ($\lambda=0.7$, $\beta=2$) to make defection-and-recovery events visible; the averaged panels use $\beta=4$.} Under high empathy, isolated defections are followed by rapid restoration of cooperation \rev{in a recovery-like sequence}, whereas low-empathy agents typically fail to recover once defection occurs. 
    \textbf{C:} Rolling action agreement rate, defined as the fraction of rounds in which both agents select the same action (either $C,C$ or $D,D$). High-empathy dyads synchronize through coordinated cooperation, low-empathy dyads synchronize primarily via mutual defection, and asymmetric empathy yields persistent desynchronization. 
    \textbf{D:} Cumulative mutual cooperation rate. \rev{High-empathy dyads converge quickly to sustained cooperative outcomes, whereas near-transition dyads accumulate cooperation more gradually and variably. Low-empathy and asymmetric interactions converge to trajectories containing few mutually cooperative outcomes.}}
    \label{fig:fig3_timeseries_abcd}
\end{figure}

 Figure~\ref{fig:fig3_timeseries_abcd}A and B illustrate these interaction trajectories. In high-empathy regimes, isolated coordination failures \rev{involving defection} are followed by rapid recovery to mutual cooperation. The \rev{rolling mutual-cooperation rate} exhibits only brief dips before returning to near unity. In contrast, low-empathy dyads display a qualitatively different pattern: \rev{defections are followed by persistent mutual defection rather than recovery}. 
    
    \par This recovery pattern can be quantified by measuring behavioral synchronization, defined as the fraction of rounds in which both agents select the same action. As shown in Figure~\ref{fig:fig3_timeseries_abcd}C, symmetric high-empathy interactions yield near-perfect synchrony, converging to coordinated cooperation within approximately ten rounds. Low-empathy dyads also synchronize, but on mutual defection. Strongly asymmetric empathy produces persistent \rev{action disagreement, reflecting systematic unilateral exploitation}. 
    
    \par Convergence to the stable regime is similarly rapid under high empathy (Figure~\ref{fig:fig3_timeseries_abcd}D). Once cooperation is established, each agent's Theory of Mind predicts continued partner cooperation, and the empathy-weighted social EFE favors maintaining $(C,C)$. Because each agent conditions its action on a history-based prediction of the opponent's behavior, mutual prediction and mutual cooperation form a self-reinforcing loop that stabilizes cooperation against \rev{isolated defections. At high symmetric empathy, a defection perturbs this history-conditioned prediction without necessarily overturning the cooperative regime, allowing the dyad to return rapidly to mutual cooperation.}
    
    \par From a dynamical systems perspective, this behavior is \rev{consistent with the emergence of a stable joint behavioral regime. When both agents place sufficiently strong and symmetric weight on the opponent's welfare, $(C,C)$ can minimize the social EFE for each simultaneously.} In this sense, coordination becomes structurally aligned: the dyad behaves as if optimizing a partially shared objective, rather than as two entirely independent payoff functions. 
    
    \par In the standard Prisoner's Dilemma, mutual defection constitutes the unique Nash equilibrium under purely self-regarding utilities \citep{kreps2018nash, nash1951,osborne1994}. \rev{Introducing empathic weighting changes each agent's effective objective and thereby alters the stability structure of the interaction: when empathic concern is sufficiently strong and symmetric, mutual cooperation becomes a stable long-run behavioral regime.}

    \subsection{Boundary-layer variability near the transition}
    \label{sec:boundary}
    
    \par
\begin{figure}[tbp]
      \centering
      \captionsetup{font=footnotesize}
    \includegraphics[width=1.0\textwidth]{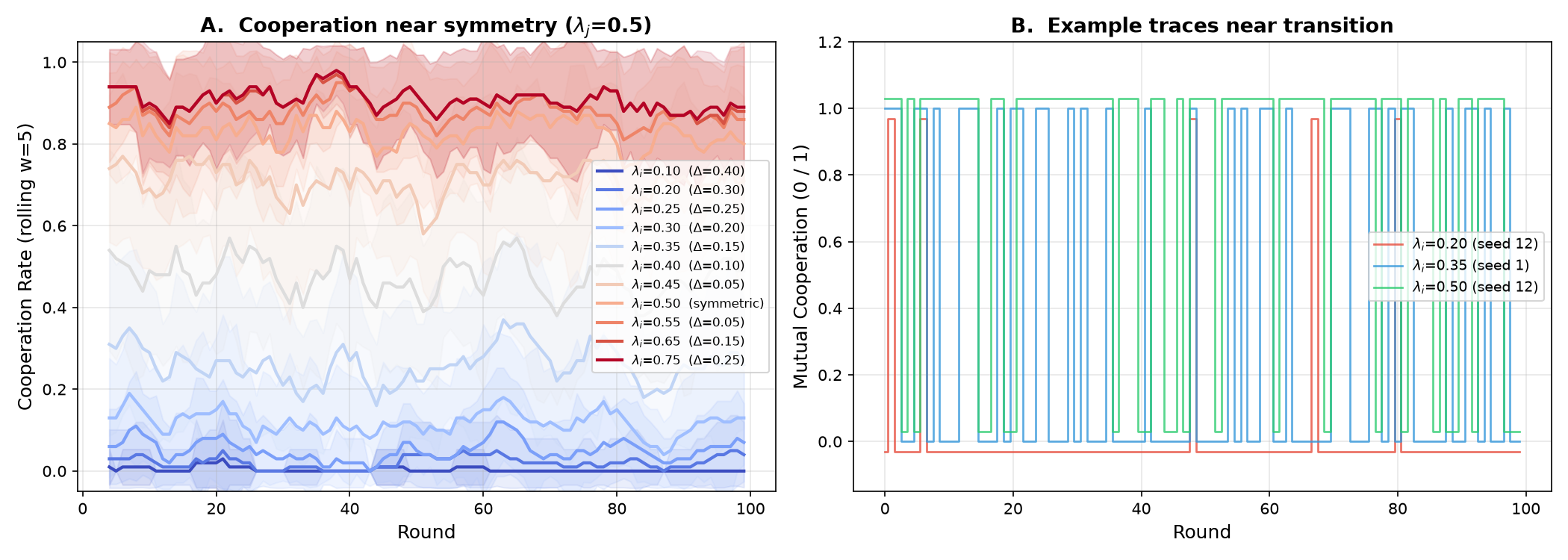}
    \includegraphics[width=1.0\textwidth]{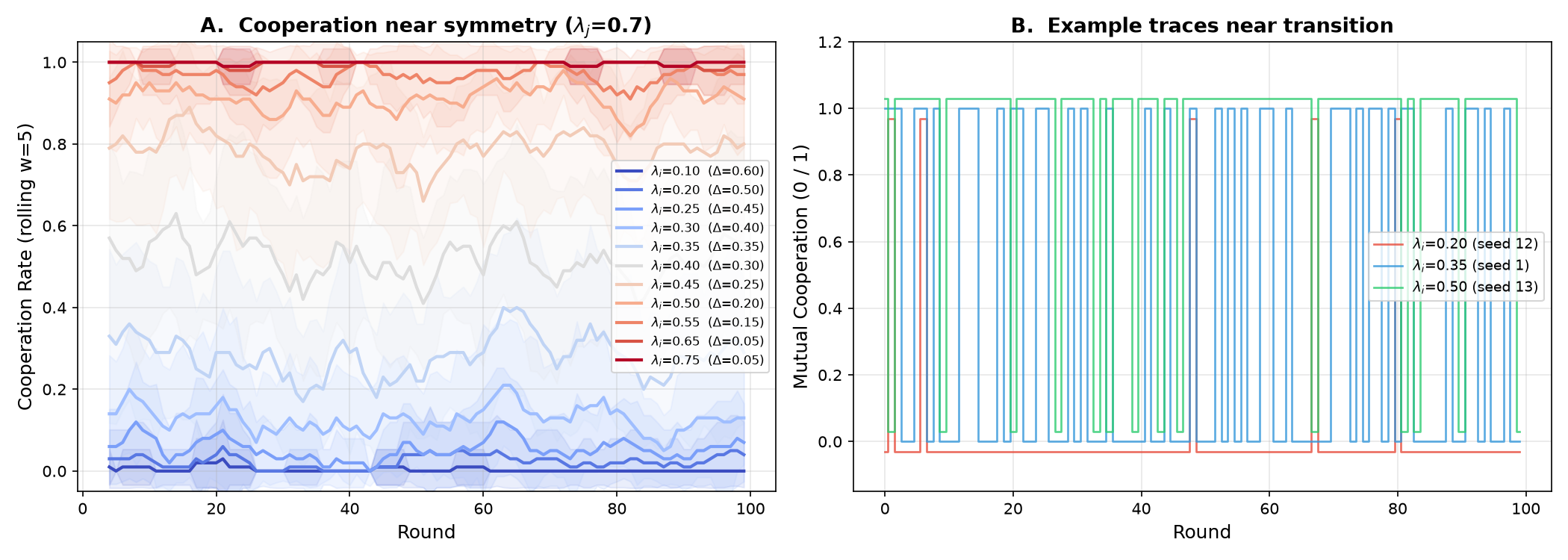}
    \caption{\textbf{\rev{Boundary-layer dynamics along fixed-partner empathy paths.}}
    \rev{All panels show 100-round interactions under myopic planning with opponent inversion disabled. The upper row fixes $\lambda_j=0.5$, and the lower row fixes $\lambda_j=0.7$. Results are averaged over 30 seeds.} 
    \textbf{A:} Rolling mutual cooperation rate (window $w=5$, mean $\pm1$ SD across seeds) as the focal agent's empathic weighting $\lambda_i$ varies. \rev{For both fixed-partner conditions, mean cooperation is approximately $0.5$ at $\lambda_i=0.20$ and rises rapidly thereafter. Within-run temporal variability peaks at this conditional crossover, reaching a mean rolling standard deviation of $0.215$ for both partner settings, before declining steadily in the post-transition cooperative region.}
    \textbf{B:} \rev{Illustrative high-variability single-seed trajectories selected within the displayed conditions. Transition-adjacent interactions show frequent switching between cooperative and non-cooperative outcomes, whereas higher values of $\lambda_i$ produce increasingly stable mutual cooperation.}}
      \label{fig:fig4_near_symmetric_dynamics}
  \end{figure}

 To characterize how the cooperation boundary depends on the dyadic empathy configuration, we fix the partner's empathic weighting at either $\lambda_j=0.5$ or $\lambda_j=0.7$ and sweep the focal agent's weighting $\lambda_i$. \rev{These fixed-partner paths are distinct from the symmetric diagonal, where $\lambda_i$ and $\lambda_j$ increase together. The analysis shown in Figure~\ref{fig:fig4_near_symmetric_dynamics} uses myopic planning ($H=1$), opponent inversion disabled, 100 interaction rounds, and 30 random seeds. Mutual cooperation rises sharply between $\lambda_i=0.20$ and $0.25$ in both partner conditions, while temporal variability is maximal at the lower edge of this conditional crossover.}

    \par To quantify the transition dynamics, we compute two variability measures for each $\lambda_i$: \rev{(1) the within-run standard deviation of the rolling mutual-cooperation rate ($w=5$), averaged across seeds, and (2) the seed-to-seed standard deviation of the interaction-level mutual-cooperation frequency, $\mathrm{freq}_{CC}$. We compare a transition-adjacent band, $\lambda_i\in\{0.20,0.25,0.30\}$, with a post-transition cooperative region, $\lambda_i\in\{0.35,0.40,0.45,0.50\}$. These regions are defined from the observed fixed-partner crossover.}

    \par \rev{Under the inversion-disabled protocol shown in Figure~\ref{fig:fig4_near_symmetric_dynamics}, mean mutual cooperation increases from $0.489$ at $\lambda_i=0.20$ to $0.724$ at $\lambda_i=0.25$ when $\lambda_j=0.5$, while the mean within-run rolling standard deviation reaches its maximum of $0.215$ at $\lambda_i=0.20$. The corresponding values for $\lambda_j=0.7$ are $0.491$, $0.727$, and $0.215$, respectively. Averaged over the transition-adjacent band, within-run variability is $0.181$ for both partner settings, compared with $0.048$ and $0.043$ in the respective post-transition regions. Seed-to-seed variability in $\mathrm{freq}_{CC}$ shows the same pattern, declining from approximately $0.039$ to $0.012$ for $\lambda_j=0.5$ and from $0.040$ to $0.010$ for $\lambda_j=0.7$.}

    \par \rev{A matched analysis with opponent inversion enabled reproduced the same qualitative pattern with similar magnitudes. The variability maximum remained at $\lambda_i=0.20$, with mean rolling standard deviations of $0.218$ and $0.217$ for $\lambda_j=0.5$ and $0.7$, respectively. In the inversion-enabled analysis, mean within-run variability decreased from $0.188$ to $0.070$ at $\lambda_j=0.5$ and from $0.187$ to $0.059$ at $\lambda_j=0.7$ between the transition-adjacent and post-transition regions (paired sign-permutation tests across 30 matched seeds, $2\times10^{4}$ permutations, both $p<10^{-4}$).}

    \par \rev{The single-seed trajectories in Figure~\ref{fig:fig4_near_symmetric_dynamics}B were deliberately selected to illustrate high-variability realizations within the displayed conditions and should not be interpreted as representative random draws. Near the conditional crossover, interactions repeatedly alternate between mutually cooperative and non-cooperative outcomes. At larger values of $\lambda_i$, these fluctuations become progressively less frequent as the dyad settles into consistently cooperative play.}
    
    \par From a dynamical perspective, these findings identify \rev{a path-dependent boundary region in the joint empathy space, rather than a universal transition determined by either agent's empathy parameter in isolation. The close agreement between the inversion-disabled and inversion-enabled sweeps further indicates that the observed boundary-layer variability is not contingent on online opponent-parameter inference.} Empathic weighting therefore influences mean cooperation as well as the temporal stability and robustness of coordination.

\par \rev{These results show that the observed location of the cooperation boundary depends on the path taken through the joint empathy space. We next isolate the symmetric diagonal, $\lambda_i=\lambda_j=\lambda$, for which a one-dimensional cooperation threshold can be defined and analytically characterized.}

\subsection{Transition to cooperation} \label{sec:transition}
\begin{figure}[tbp]
    \centering
    \captionsetup{font=footnotesize}
    \includegraphics[width=0.95\textwidth]{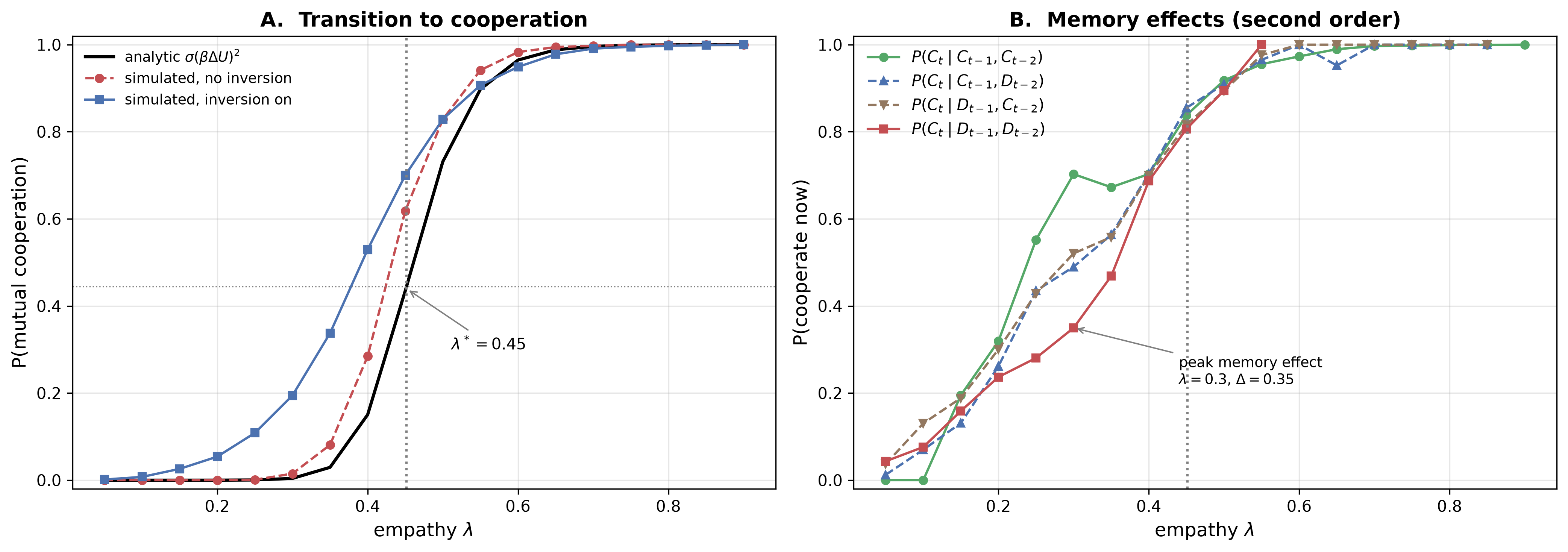}
    \caption{\textbf{\rev{Transition to cooperation along the symmetric empathy diagonal.}}
    \textbf{A:} \rev{Mutual cooperation probability as a function of the shared empathic weighting $\lambda$, with $\lambda_i=\lambda_j=\lambda$. The analytical curve is derived for identical agents without learning or opponent inversion, and the corresponding inversion-disabled simulation provides the like-for-like empirical comparison. The inversion-enabled curve shows the full history-conditioned opponent model. The analytical point of maximal sensitivity is $\lambda^*=0.451$, whereas the deterministic indifference point is $\lambda_0=0.420$.}
    \textbf{B:} \rev{History-conditioned cooperation probabilities under opponent inversion, measured by conditioning the agent's current action on its own previous two actions. At $\lambda=0.35$, the probability of cooperation decreases from $0.673$ following two cooperations to $0.469$ following two defections. The separation among the four conditionals is maximal near $\lambda=0.30$, where it reaches approximately $0.35$, and narrows as cooperation approaches saturation. Disabling opponent inversion limits the corresponding spread to at most approximately $0.07$, indicating that this apparent memory structure is mediated by the history-conditioned opponent model rather than by an explicit second-order memory term in the action rule.}}
    \label{fig:fig5_pcc}
\end{figure}

\par \rev{Having characterized conditional crossovers along fixed-partner paths in Figure~\ref{fig:fig4_near_symmetric_dynamics}, we now isolate the symmetric diagonal of the joint empathy space, $\lambda_i=\lambda_j=\lambda$. This restriction yields a one-dimensional control parameter for which the empirical cooperation curve can be compared directly with an analytical approximation. Unless otherwise stated, the empirical curves in Figure~\ref{fig:fig5_pcc} summarize 100-round interactions over 20 random seeds under myopic planning and fixed empathic parameters.}

\par \rev{Mutual cooperation increases nonlinearly along the symmetric diagonal. With opponent inversion enabled, $P(CC)$ rises from $0.195$ at $\lambda=0.30$ to $0.530$ at $\lambda=0.40$, $0.829$ at $\lambda=0.50$, and $0.991$ at $\lambda=0.70$. Without opponent inversion, the corresponding values are $0.015$, $0.285$, $0.829$, and $0.997$. Opponent inversion therefore shifts the lower portion of the empirical crossover toward weaker empathic weighting, while both conditions converge to nearly complete cooperation at sufficiently high $\lambda$. On the sampled grid, the inversion-enabled curve first exceeds $P(CC)=0.5$ at $\lambda=0.40$ and exceeds $P(CC)=0.8$ at $\lambda=0.50$; these operational cutoffs should be distinguished from the analytical maximal-sensitivity point defined below.}

\par To analytically characterize the symmetric crossover, we consider identical agents without learning or opponent inversion. For agent $i$, define the empathy-weighted expected utility of action $a_i$ as:
\begin{equation}
    U(a_i,\lambda) = \sum_{a_j} q(a_j\mid a_i)
    \left[ (1-\lambda)u_i(a_i,a_j) + \lambda u_j(a_i,a_j) \right]
\end{equation}

\par Here, the utility difference between cooperation and defection is: 
\begin{equation}
    \Delta U(\lambda) = U(C,\lambda)-U(D,\lambda)
\end{equation}

\par And the probability of cooperation is approximated by: 
\begin{equation} 
    P_i(C) = \sigma\!\left(\beta\Delta U(\lambda)\right),
    \qquad \sigma(x)=\frac{1}{1+\exp(-x)}
\end{equation}

\par \rev{Under the additional approximation that the two identical agents' actions are conditionally independent,}
\begin{equation}
    P(CC) \approx P_i(C)P_j(C) = p^2(\lambda,\beta)
\end{equation}

\par \rev{This closed-form approximation neglects the higher-order temporal coupling produced by repeated interaction and is therefore compared most directly with the inversion-disabled simulation rather than treated as an exact description of the full dyad.}

\par We define the analytical transition point $\lambda^*$ as the value at which $P(CC)$ is maximally sensitive to changes in empathic weighting. For the monotonic finite-precision crossover considered here, this corresponds to the unique interior point satisfying $\left. \frac{d^2P(CC)}{d\lambda^2} \right|_{\lambda=\lambda^*} = 0$.

\par Under the assumptions above, with $P(CC)=p^2(\lambda,\beta)$ and $p(\lambda,\beta)=\sigma\!\left(\beta[A+\lambda B]\right)$, this condition gives $p(\lambda^*)=2/3$, and
\begin{equation}
    \lambda^* = \frac{\ln 2/\beta-A}{B}
\end{equation}
where $A=q_0R+(1-q_0)S-q_1T-(1-q_1)P$, $B=(T-S)(1-q_0+q_1)$, and $q_0\equiv q(a_j=C\mid a_i=C)$, $q_1\equiv q(a_j=C\mid a_i=D)$.

\par \rev{At $\lambda^*$, the individual cooperation probability is $p(\lambda^*)=2/3$, and hence the independence approximation gives $P(CC)=4/9$. For the opponent model used here, $q_0=0.881$, $q_1=0.998$, $A=-2.348$, and $B=5.584$, yielding $\lambda^*=0.451$. The inversion-disabled simulation exhibits similar crossover in the same parameter region. In the deterministic limit, $\beta\rightarrow\infty$, the maximal-sensitivity point approaches the utility-indifference value $\lambda_0 = -\frac{A}{B} = 0.420$.} \rev{Note that the inversion-enabled empirical curve is not expected to share the same closed-form transition point because history-conditioned opponent inference introduces additional coupling; therefore, its sampled operational thresholds are reported separately rather than denoted by the same $\lambda^*$.}

\par \rev{Figure~\ref{fig:fig5_pcc}B reveals a related history-conditioned signature on the lower shoulder of the crossover. At $\lambda=0.10$, cooperation remains effectively absent even following two previous cooperative actions, with $P(C_t\mid C_{t-1},C_{t-2})=0.000$. At $\lambda=0.35$, cooperation is more fragile: its probability falls from $0.673$ following two cooperations to $0.469$ following two defections. The maximal separation among the four second-order conditionals occurs near $\lambda=0.30$, where it reaches approximately $0.35$, and the conditionals generally converge as mutual cooperation saturates.}

\par \rev{These conditional differences should be interpreted as an emergent history dependence of the coupled interaction rather than as an explicit second-order memory mechanism. When opponent inversion is disabled, the four conditionals remain within approximately $0.07$ of one another across the tested range. The pronounced separation under inversion therefore reflects how recent action history reshapes the agent's prediction of its partner, which then enters social EFE evaluation. Together, the analytical and empirical results characterize the symmetric transition as a smooth but sharp finite-precision crossover whose location and temporal structure are modified by opponent inference.}

\subsection{Learning \rev{refines opponent beliefs but does not substitute for prosocial valuation}}
\label{sec:learning}

\par To disentangle the role of \rev{adaptive opponent inference} from that of social valuation, we evaluate learning-enabled variants in which each agent maintains a particle filter posterior over continuous opponent parameters $\theta_j = (\alpha, \rho, \beta, \lambda_j)$, representing cooperation bias, reciprocity sensitivity, behavioral precision, and empathic weighting, respectively. Rather than inferring discrete strategy types, the agent performs online Bayesian inference over this parametric behavioral profile.

\par
\begin{figure}[tbp]
    \centering
    \captionsetup{font=footnotesize}
    \includegraphics[width=1.0\textwidth]{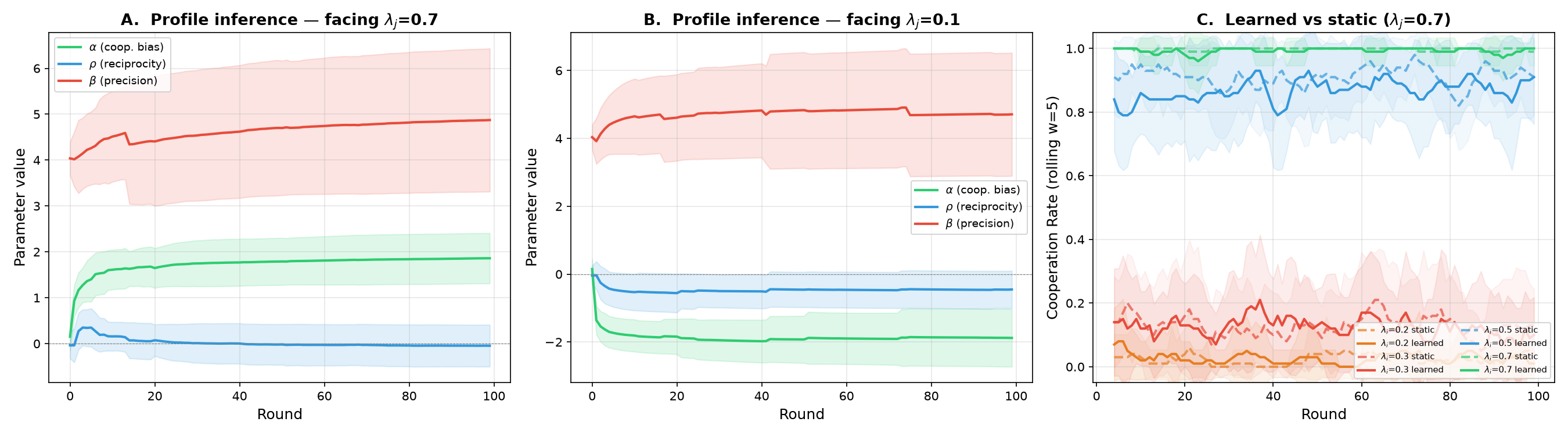}
    \caption{\textbf{Opponent parameter inference and its effect on cooperation.}
    \textbf{A:} Posterior mean and uncertainty (shaded bands) over opponent behavioral parameters $(\alpha,\rho,\beta)$ when facing a cooperative partner ($\lambda_j = 0.7$). \rev{The inferred cooperation bias $\alpha$ converges to a strongly positive value, while posterior uncertainty contracts over repeated interaction.}
    \textbf{B:} Parameter inference when facing a low-empathy partner ($\lambda_j = 0.1$). The inferred cooperation bias $\alpha$ converges to a negative value, consistent with sustained defection. In both partner conditions, posterior \rev{estimates stabilize and uncertainty decreases over time.}
    \textbf{C:} Mutual cooperation rate under learning-enabled inference (solid lines) versus static Theory of Mind (dashed lines), with the partner fixed at $\lambda_j=0.7$. \rev{Learning produces a small increase in cooperation, most visibly near the conditional boundary at $\lambda_i=0.4$, but does not move weakly empathic focal agents into a robust cooperative regime. At higher focal empathy, cooperation remains near ceiling regardless of whether online learning is enabled.}}
    \label{fig:fig6_learning_convergence}
\end{figure}

 Figure~\ref{fig:fig6_learning_convergence}A and B show the evolution of three inferred behavioral parameters, $(\alpha,\rho,\beta)$, over repeated interaction. When facing a cooperative partner ($\lambda_j = 0.7$), the posterior mean of $\alpha$ converges to a strongly positive value, \rev{consistent with persistent cooperative behavior,} while reciprocity $\rho$ and precision $\beta$ stabilize \rev{as posterior uncertainty contracts}. When facing a low-empathy partner ($\lambda_j = 0.1$), the inferred cooperation bias $\alpha$ converges to a negative value, reflecting sustained defection, and uncertainty again decreases over time. \rev{These trajectories indicate stable posterior contraction and differentiation between cooperative and defective opponent profiles.}

\par \rev{Improved opponent modelling modestly increases cooperation but does not, on its own, establish a robust cooperative regime.} Figure~\ref{fig:fig6_learning_convergence}C compares learning-enabled inference with static ToM while fixing the partner at $\lambda_j=0.7$ and varying the focal agent's empathic weighting $\lambda_i$. \rev{At $\lambda_i=0.2$, mutual cooperation increases from $1.8\%$ under static ToM to $2.2\%$ with learning enabled, and at $\lambda_i=0.3$ it increases from $11.7\%$ to $13.7\%$. The largest observed change occurs at $\lambda_i=0.4$, where learning raises mutual cooperation from $46.9\%$ to $52.8\%$, a gain of $5.8$ percentage points. Thus, recognizing a cooperative partner can facilitate cooperation near the boundary, but the effect remains small relative to that of the focal agent's empathic weighting. At $\lambda_i\geq0.5$, cooperation is already near ceiling under both inference conditions.}

\par \rev{These results distinguish the behavioral contribution of opponent inference from that of other-regarding valuation. Learning updates and sharpens the agent's posterior representation of its partner and can produce a modest upward shift in cooperation when the partner is cooperative. It does not, however, substitute for empathic weighting in the social EFE: focal agents with weak other-regarding valuation remain predominantly non-cooperative despite learning that the partner is cooperative. Cooperation is therefore governed primarily by how inferred partner behavior is valued under $\lambda$, rather than by opponent identification alone. The direct model comparison in Section~\ref{sec:model_comparison} tests this dissociation more stringently by holding opponent-prediction machinery constant while varying other-regarding valuation.}

\subsection{Model comparison: dissociating prediction from valuation}
\label{sec:model_comparison}

\par \rev{The preceding learning analysis shows that improved opponent inference can modestly facilitate cooperation but does not substitute for other-regarding valuation. To isolate these contributions directly, we compare three agent types that share the architecture of Section~\ref{sec:methodology} and differ only in empathic weighting and whether opponent inversion is active: a \emph{self-interested} agent ($\lambda=0$, inversion off), a \emph{ToM-only} agent ($\lambda=0$, inversion on), and an \emph{empathic} agent ($\lambda=0.6$, inversion on). The comparison between the self-interested and ToM-only agents isolates the contribution of opponent-sensitive prediction at fixed valuation, whereas the comparison between the ToM-only and empathic agents isolates the contribution of other-regarding valuation while retaining the same inversion machinery. All pairings were run for $T=100$ rounds under myopic planning, with 25 seeds per cell.}

\par
\begin{figure}[tbp]
    \centering
    \captionsetup{font=footnotesize}
    \includegraphics[width=1.0\textwidth]{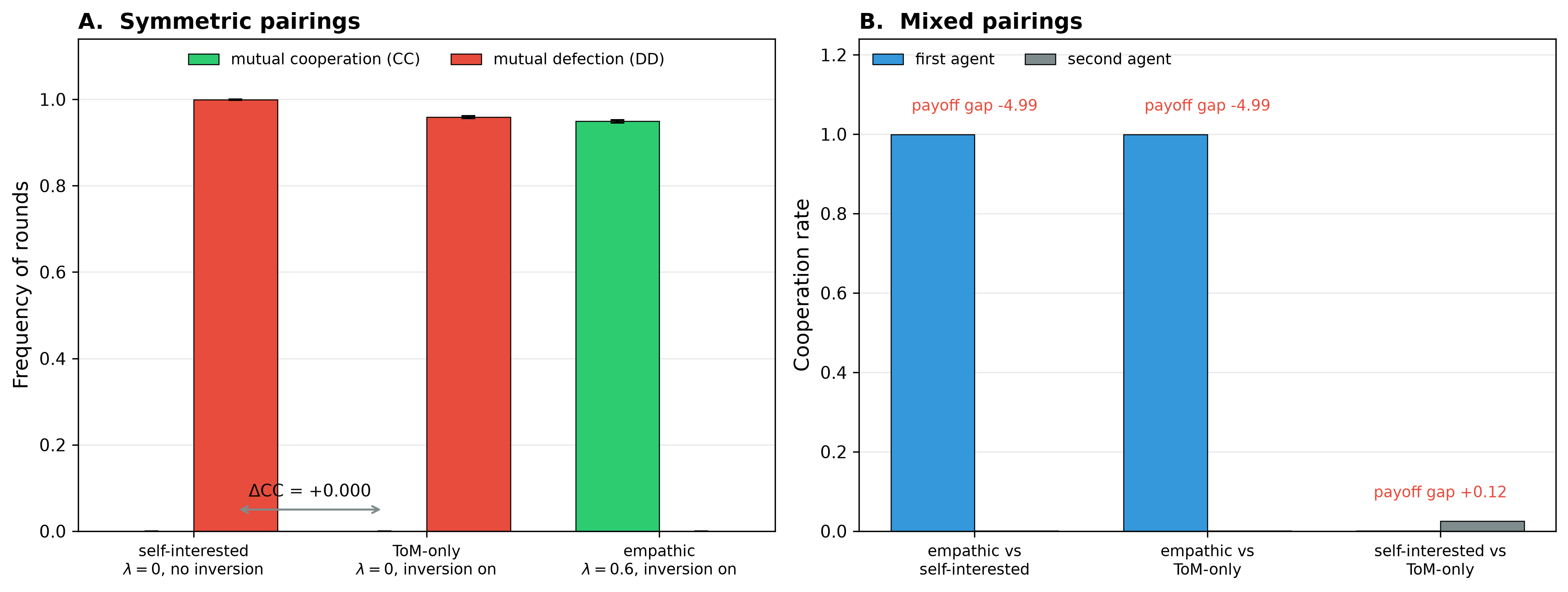}
    \caption{\rev{\textbf{Prediction and valuation are dissociable.}
    \textbf{A:} Symmetric pairings. Self-interested and ToM-only dyads exhibit effectively zero mutual cooperation; adding opponent inversion at $\lambda=0$ leaves mean mutual cooperation unchanged ($0.000$). Empathic dyads, using the same inversion machinery as the ToM-only agents but with $\lambda=0.6$, converge to mutual cooperation. Error bars denote $\pm 1$ s.e.m. across 25 seeds.
    \textbf{B:} Mixed pairings, showing each agent's own cooperation rate. The empathic agent continues to cooperate while either non-empathic partner defects, incurring a mean payoff disadvantage of approximately $4.9$ units per round. The self-interested versus ToM-only pairing remains approximately symmetric.}}
    \label{fig:fig8_model_comparison}
\end{figure}

 \rev{Table~\ref{tab:model_comparison} and Figure~\ref{fig:fig8_model_comparison} report the outcomes. In symmetric pairings, self-interested and ToM-only dyads are behaviorally nearly indistinguishable: both exhibit effectively zero mutual cooperation, and the ToM-only minus self-interested difference in mean mutual cooperation is $0.000$. Adding opponent inversion without other-regarding valuation therefore does not generate cooperation. By contrast, empathic dyads, which use the same inversion machinery as the ToM-only agents but assign positive weight to the partner's welfare, converge to sustained mutual cooperation.}

\par \rev{Mixed pairings expose the corresponding cost of fixed empathic valuation. An empathic agent paired with either non-empathic type continues to cooperate while its partner defects, yielding exploitability of $1.00$ and mean per-round payoff gaps of $-4.99$ against both the self-interested and the ToM-only agent, measured from the empathic agent's perspective. The self-interested versus ToM-only pairing remains approximately symmetric. Together, these comparisons establish the dissociation: opponent-sensitive prediction without other-regarding valuation does not produce cooperation, whereas changing valuation while retaining the same prediction machinery does. Prediction can inform action selection, but whether the predicted outcome is pursued depends on how the partner's welfare enters the agent's objective.}

\begin{table}[tbp]
  \centering
  \captionsetup{font=footnotesize}
  \setlength{\tabcolsep}{8pt}
  \caption{\rev{Model comparison ($T=100$, 25 seeds, myopic planning). The upper block reports mutual-cooperation and mutual-defection frequencies for symmetric pairings. The lower block reports mixed pairings; exploitability is the absolute difference between the agents' cooperation rates, and the payoff gap is the first-listed agent's mean per-round payoff minus that of its partner.}}
  \label{tab:model_comparison}
  \begin{tabular}{lcc}
    \hline
    Pairing & CC freq. & DD freq. \\
    \hline
    self-interested vs self-interested & 0.00 & 1.00 \\
    ToM-only vs ToM-only & 0.00 & 0.96 \\
    empathic vs empathic & 0.95 & 0.00 \\
    \hline
     & Exploitability & Payoff gap \\
    \hline
    self-interested vs ToM-only & 0.02 & $+0.12$ \\
    empathic vs self-interested & 1.00 & $-4.99$ \\
    empathic vs ToM-only & 1.00 & $-4.99$ \\
    \hline
  \end{tabular}
\end{table}

\par \rev{To clarify the representational scope of the scalar empathy parameter, we separately examine the sign-extended self/other weight space introduced in Section~\ref{sec:methodology}. On the standard simplex, $w_{\mathrm{self}}=1-\lambda$ and $w_{\mathrm{other}}=\lambda$, so both weights are non-negative. Allowing either weight to become negative introduces regimes that cannot be represented by $\lambda\in[0,1]$.}

\par \rev{Negative other-regarding weight does not produce a behaviorally distinct spite regime under the present payoff structure. Symmetric dyads at $(w_{\mathrm{self}},w_{\mathrm{other}})=(0.6,-0.4)$ and under pure spite $(0,-1)$ both reach mutual defection on $100.0\%$ of rounds, compared with $99.9\%$ for the purely self-interested condition $(1,0)$. Thus, valuing the partner's loss and assigning no value to the partner's welfare lead to practically indistinguishable behavior in this game. By contrast, the tested self-abnegating condition, $(w_{\mathrm{self}},w_{\mathrm{other}})=(-0.4,0.6)$, produces mutual cooperation on $100.0\%$ of rounds, compared with $98.3\%$ at the most cooperative point on the non-negative simplex. The sign extension therefore reveals a distinct self-abnegating regime, but not a distinct behavioral effect of spite in the present IPD configuration.}

\begin{figure}[tbp]
    \centering
    \captionsetup{font=footnotesize}
    \includegraphics[width=1.0\textwidth]{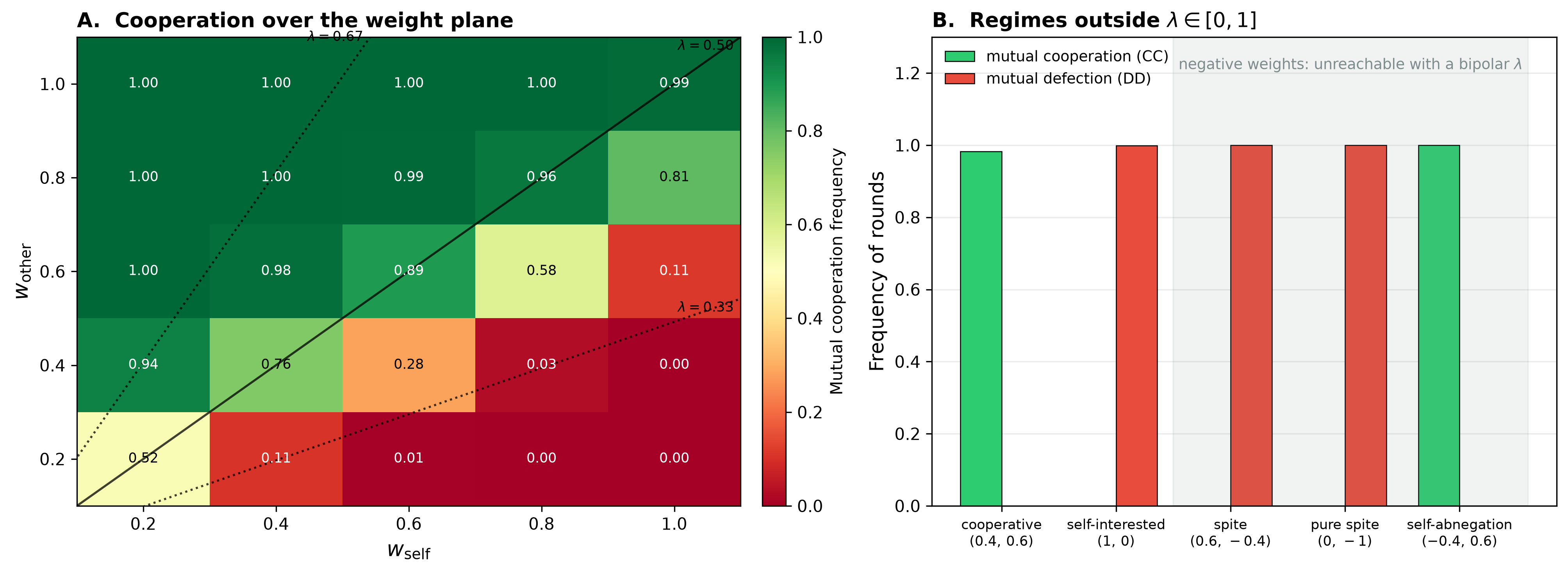}
    \caption{\rev{\textbf{The self/other weight space.}
    \textbf{A:} Mutual cooperation frequency over the positive quadrant of $(w_{\mathrm{self}},w_{\mathrm{other}})$, with 20 seeds per cell. The direction of a ray through the origin determines the relative other-regarding weight $\lambda=w_{\mathrm{other}}/(w_{\mathrm{self}}+w_{\mathrm{other}})$, whereas the tested self-abnegating condition with negative $w_{\mathrm{self}}$ and positive $w_{\mathrm{other}}$ produces complete mutual cooperation. 
    \textbf{B:} Sign-extended regimes that cannot be represented by the non-negative simplex $\lambda\in[0,1]$. Negative $w_{\mathrm{other}}$ produces behavior that is effectively indistinguishable from pure self-interest under the present payoff structure, whereas negative $w_{\mathrm{self}}$ with positive $w_{\mathrm{other}}$ produces complete mutual cooperation.}}
    \label{fig:fig10_weight_space}
\end{figure}

\rev{\subsection{Planning depth amplifies the opponent model}}
\label{sec:sophistication}

\par All results presented thus far were obtained under myopic action selection ($H = 1$), in which agents evaluate only the immediate social EFE of each candidate action. We now ask how multi-step planning changes cooperation and whether its effect depends on the predicted responsiveness of the partner.

\par To this end, we compare myopic agents with sophisticated agents that enumerate all $2^H$ candidate policies over planning horizons $H = 2$ and $H = 3$, evaluating each policy through forward rollout (described in Section~\ref{sec:methodology}). \rev{In the symmetric-empathy mixed-horizon analysis, the sophisticated agent is paired with a partner that shares the same empathy parameter but plans myopically ($H=1$); the table reports the dyad's mutual-cooperation frequency. At rollout step $t=0$, the partner's predicted response is obtained from the current gated, history-conditioned Theory of Mind model. At $t>0$, the prediction is conditioned on the simulated history generated by the candidate policy: the focal agent's simulated previous action enters the opponent model's reciprocity term, and its simulated cooperation rate enters the opponent's belief about the focal policy. Without this prefix conditioning, the predicted response would be identical across candidate policies and the horizon would not affect action selection.} 

\par The policy-level social EFE is \rev{accumulated} across the horizon as: 
\begin{equation}
    G(\pi) = \sum_{t=0}^{H-1}
    \left[
    (1-\lambda)G_{\text{self}}^{(t)}
    + \lambda G_{\text{other}}^{(t)}
    + G_{\text{epistemic}}^{(t)}
    \right]
\end{equation}
\par Actions are selected by marginalizing the softmax posterior $Q(\pi)\propto\exp[-\beta G(\pi)]$ to the first time step. The horizon sum is not normalized by $H$, and all other parameters are held constant.

\begin{table*}[tbp]
  \centering
  \captionsetup{font=footnotesize}
  \caption{\rev{Mutual cooperation frequency by planning horizon in symmetric-empathy mixed-horizon dyads. A sophisticated focal agent is paired with a myopic partner of equal empathy ($T=100$, 20 seeds, opponent inversion enabled). The horizon sum is unnormalised, and rollout predictions are conditioned on the simulated prefix; see Section~\ref{sec:methodology}.}}
  \label{tab:tab2_sophisticated}
  \setlength{\tabcolsep}{8pt}
  \begin{tabular}{c c c c c}
  \hline
  $\lambda$ & $H = 1$ & $H = 2$ & $H = 3$ & $\Delta$(H3$-$H1) \\
  \hline
  0.1 & 0.007 & 0.006 & 0.007 & $+0.000$ \\
  0.2 & 0.054 & 0.036 & 0.035 & $-0.019$ \\
  0.3 & 0.195 & 0.150 & 0.141 & $-0.054$ \\
  0.4 & 0.529 & 0.478 & 0.461 & $-0.068$ \\
  0.5 & 0.829 & 0.836 & 0.836 & $+0.007$ \\
  0.7 & 0.991 & 0.993 & 0.993 & $+0.002$ \\
  \hline
  \end{tabular}
\end{table*}

\begin{figure}[tbp]
      \centering
      \captionsetup{font=footnotesize}
      \includegraphics[width=1.0\textwidth]{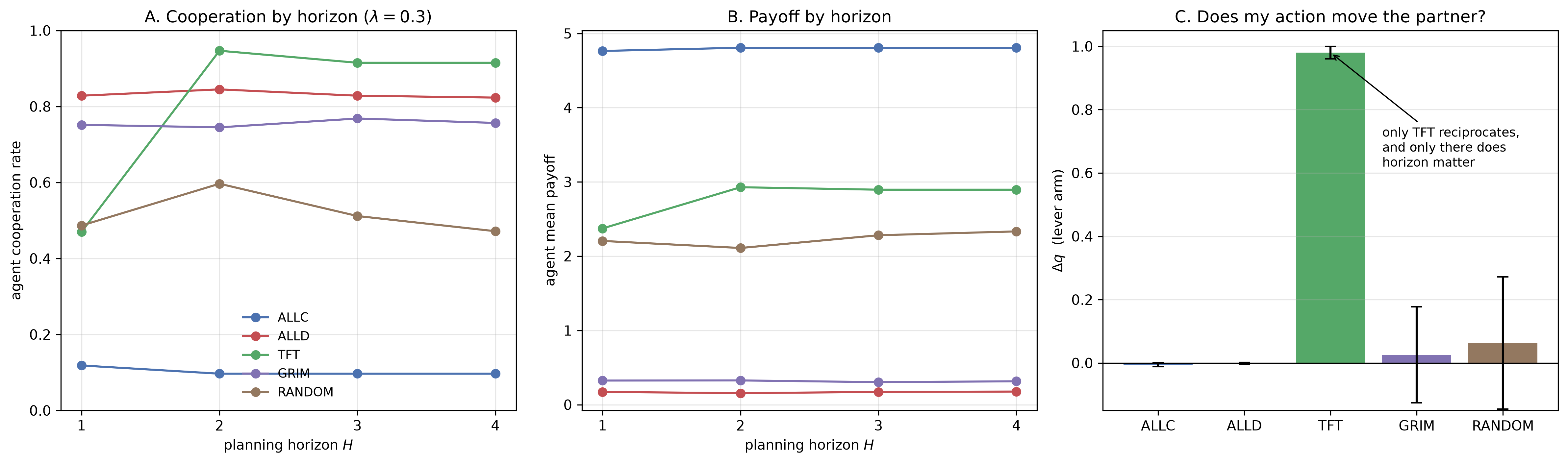}
      \caption{\rev{\textbf{Planning depth amplifies predicted partner responsiveness.}
      \textbf{A:} Agent cooperation rate against each fixed partner strategy as a function of planning horizon $H$ ($\lambda=0.3$, $T=60$, 10 seeds). Among the tested partners, tit-for-tat is the only strategy for which planning depth produces a large behavioral change, raising cooperation from $0.470$ at $H=1$ to $0.947$ at $H=2$.
      \textbf{B:} Agent mean payoff for the same conditions; against tit-for-tat, the increase from $H=1$ to $H=2$ is $+0.522$.
      \textbf{C:} The model-based responsiveness measure $\Delta q = q(a_j=C \mid a_i^{t-1}=C) - q(a_j=C \mid a_i^{t-1}=D)$ after 60 rounds, with standard deviations across seeds. Tit-for-tat is the only tested partner for which the focal agent's previous action strongly changes the predicted response.}}
      \label{fig:fig7_partner_planning}
\end{figure}

\par Table~\ref{tab:tab2_sophisticated} shows that deeper planning produces \rev{small negative shifts in mutual cooperation on the lower shoulder of the symmetric crossover}. At $\lambda=0.3$, increasing the horizon from $H=1$ to $H=3$ reduces mutual cooperation from $0.195$ to $0.141$ ($\Delta CC=-0.054$); the largest reduction on the tested grid occurs at $\lambda=0.4$ ($\Delta CC=-0.068$). At $\lambda\geq0.5$, the corresponding changes are numerically small ($+0.007$ and $+0.002$). \rev{These values rely on an unnormalized horizon sum. Dividing the policy-level EFE by $H$ while holding $\beta$ fixed would simultaneously reduce effective action precision and therefore confound foresight with increased decision noise.} \rev{The self-play effect is concentrated near the symmetric cooperation crossover identified in Section~\ref{sec:transition}. At $\lambda=0.1$ and $0.2$, cooperation is already rare at every horizon, whereas at $\lambda\geq0.5$ it is near the ceiling. Planning has its largest effect at $\lambda=0.3$ and $0.4$, where cooperation is feasible but not yet saturated.}

\par \rev{Mechanistically, planning can change the first action only when different simulated actions lead the model to predict different future partner responses. For fixed partner strategies, this dependence is summarized by $\Delta q = q(a_j=C \mid a_i^{t-1}=C) - q(a_j=C \mid a_i^{t-1}=D)$. Tit-for-tat produces a large positive lever arm, $\Delta q\approx0.99$, whereas the other tested partners yield values between approximately $-0.005$ and $0.08$. In symmetric self-play, the corresponding realized behavioral responsiveness is also weak: the lag-1 correlation between one agent's action and its partner's next action is $-0.076$, $-0.033$, and $-0.010$ at $\lambda=0.3$, $0.5$, and $0.7$, respectively.}

\par \rev{This distinction explains the reversal across partner types. In self-play, cooperating at the current round buys little additional partner cooperation later, so deeper planning slightly favors defection-initial policies. Against tit-for-tat, by contrast, current cooperation strongly increases the predicted probability of future partner cooperation. Increasing the horizon from $H=1$ to $H=2$ therefore raises the focal agent's cooperation rate from $0.470$ to $0.947$ and its mean payoff from $2.373$ to $2.928$. Planning depth does not intrinsically favor cooperation or defection; instead, it magnifies the consequences encoded by the opponent model.}

\par \rev{We also define an operational high-cooperation threshold $\lambda_{0.8} = \min\{\lambda:P(CC)\geq0.8\}$. On the grid used in Table~\ref{tab:tab2_sophisticated}, $\lambda_{0.8}=0.50$ at every tested horizon. Thus, although planning slightly changes cooperation within the crossover region, it does not measurably shift the empathic weighting required to exceed $80\%$ mutual cooperation. We denote this operational cutoff by $\lambda_{0.8}$ to distinguish it from the analytical maximal-sensitivity point $\lambda^*$ defined in Section~\ref{sec:transition}.}

\par \rev{The modest reduction observed in self-play is reminiscent of the strategic pressure toward defection emphasized by backward-induction analyses of the finitely repeated Prisoner's Dilemma \citep{osborne1994}. However, the reversal against tit-for-tat shows that foresight is not intrinsically anti-cooperative in the present model. Its effect depends on whether the predicted partner dynamics reward current cooperation with future cooperation. These results sharpen the distinction between planning capability and social valuation. Increasing planning depth does not itself add prosocial concern; instead, it makes the agent more sensitive to the modeled long-run consequences of its actions. When the opponent model predicts weak or negative reciprocity, this can modestly reduce cooperation. When it predicts strong positive reciprocity, the same planning mechanism can produce a large cooperative gain. Alignment in this framework therefore depends jointly on the agent's other-regarding valuation and on the interaction structure represented by its opponent model.}

\par \rev{To examine whether the payoff dependence of the rollout planner is specific to the standard Prisoner's Dilemma matrix, we conduct a complementary sensitivity analysis in a reduced static-ToM setting. We consider three payoff structures satisfying $T>R>P>S$ and $2R>T+S$: the standard matrix $(R,S,T,P)=(3,0,5,1)$, a weak-temptation matrix $(3,1,4,2)$, and a high-temptation matrix $(5,0,8,1)$. This analysis disables opponent inversion and evaluates the partner under the fixed static-ToM baseline, rather than the history-conditioned opponent model used in the main mixed-horizon analysis. Simulations use $T=100$ rounds, 20 seeds per cell, and horizons $H\in\{1,2,3,4\}$. The analysis therefore probes the payoff sensitivity of the planner itself; its absolute cooperation levels are not directly comparable with Table~\ref{tab:tab2_sophisticated}.}

\par \rev{Within this reduced static-ToM setting, deeper planning lowers cooperation in every tested condition for which the myopic baseline is already strongly cooperative. The pattern occurs under all three payoff matrices, although its magnitude depends on the payoff-empathy combination. Its direction is not universal: under weak temptation at $\lambda=0.3$, where myopic cooperation is only $0.17$, cooperation instead rises slightly to $0.20$ at $H=4$. Thus, the rollout planner amplifies the action preference induced jointly by empathic weighting and the payoff margins, rather than imposing a uniformly cooperative or defective horizon effect. The numerical coincidence between the standard-payoff row at $\lambda=0.3$ and the weak-temptation row at $\lambda=0.5$ further illustrates this shared dependence. Because this analysis uses a reduced opponent model, extension of the same payoff comparison to the full inversion-enabled model remains a target for future work.}

\begin{table}[tbp]
  \centering
  \captionsetup{font=footnotesize}
  \setlength{\tabcolsep}{8pt}
  \caption{\rev{Payoff sensitivity of the rollout planner in the reduced static-ToM setting. Mutual-cooperation frequency by planning horizon and payoff structure ($T=100$, 20 seeds). Note that the opponent inversion is disabled and the partner is evaluated under the fixed static-ToM baseline; absolute cooperation levels are therefore not directly comparable with Table~\ref{tab:tab2_sophisticated}.}}
  \label{tab:payoff_horizon}
  \begin{tabular}{llcccc}
    \hline
    Payoffs $(R,S,T,P)$ & $\lambda$ & $H=1$ & $H=2$ & $H=3$ & $H=4$ \\
    \hline
    standard $(3,0,5,1)$ & 0.3 & 0.78 & 0.66 & 0.60 & 0.56 \\
     & 0.5 & 1.00 & 0.95 & 0.89 & 0.83 \\
     & 0.7 & 1.00 & 0.99 & 0.97 & 0.92 \\
    \hline
    weak $(3,1,4,2)$ & 0.3 & 0.17 & 0.19 & 0.19 & 0.20 \\
     & 0.5 & 0.78 & 0.66 & 0.60 & 0.56 \\
     & 0.7 & 0.98 & 0.89 & 0.81 & 0.76 \\
    \hline
    high $(5,0,8,1)$ & 0.3 & 0.99 & 0.94 & 0.87 & 0.81 \\
     & 0.5 & 1.00 & 1.00 & 0.98 & 0.95 \\
     & 0.7 & 1.00 & 1.00 & 1.00 & 0.99 \\
    \hline
  \end{tabular}
\end{table}

\begin{figure}[tbp]
    \centering
    \captionsetup{font=footnotesize}
    \includegraphics[width=1.0\textwidth]{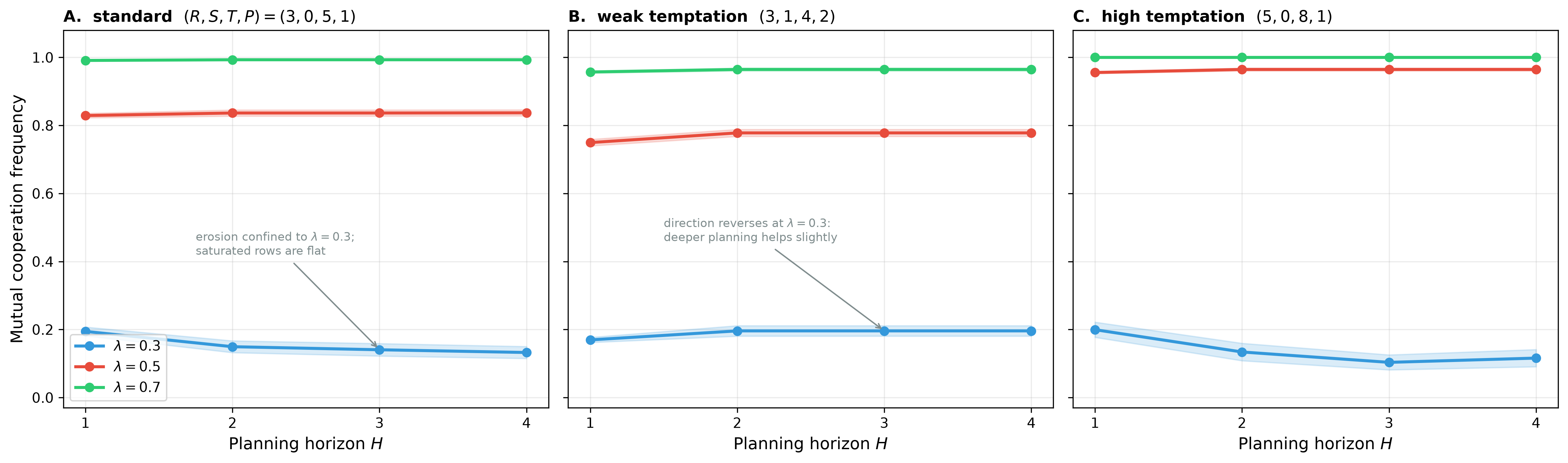}
    \caption{\rev{\textbf{Payoff sensitivity of planning depth near the cooperation transition.} Mutual-cooperation frequency as a function of planning horizon $H$ under the standard payoffs (\textbf{A}), weak temptation (\textbf{B}), and high temptation (\textbf{C}). All three matrices satisfy $T>R>P>S$ and $2R>T+S$. Shaded bands denote $\pm1$ s.e.m. across 20 seeds. Both agents use opponent inversion, as elsewhere in the paper. Planning depth matters only near the transition: at $\lambda=0.3$ it erodes cooperation under the standard and high-temptation payoffs and slightly increases it under weak temptation, whereas at $\lambda=0.5$ and $\lambda=0.7$ cooperation is saturated and horizon has no discernible effect. Under a static opponent model these two curves were degenerate: empathic weighting and payoff margins entered the social EFE through the same weighted combination. Inferring $\lambda_j$ lifts that degeneracy, so payoff structure and empathic weight are no longer interchangeable.}}
    \label{fig:fig9_payoff_horizon}
\end{figure}

\section{Discussion}\label{sec:discussion}

\par \rev{The present results identify other-regarding valuation as a distinct control dimension in repeated social interaction. Increasing empathic weighting reorganizes the dyadic cooperation landscape, but its effects depend on the joint configuration of both agents: sufficiently symmetric weighting supports sustained mutual cooperation, whereas strong asymmetry exposes the more empathic agent to systematic exploitation. The resulting cooperation boundary is path-dependent, with elevated temporal variability along fixed-partner crossings and a sharp but continuous crossover along the symmetric diagonal. Online opponent learning modestly facilitates cooperation near this boundary but does not substitute for other-regarding valuation. Likewise, planning depth has no intrinsically cooperative or defective effect; it amplifies the future interaction structure represented by the opponent model, slightly reducing cooperation in weakly reciprocal self-play while strongly increasing it against a reciprocating partner.}

  \subsection{Relation to active inference literature}

  The proposed empathy-driven cooperation mechanism builds on multiple strands of work in the active inference literature, particularly research on social interaction, games, and multi-agent coordination. One important foundation is active inference applied to strategic games and social dilemmas. \citet{demekas2023analytical} introduced a mathematically tractable active inference formulation of the Iterated Prisoner's Dilemma, demonstrating how agents can learn adaptive responses over repeated interactions. In that formulation, however, agents reason over the joint state space of the game, enumerating possible action combinations, without maintaining distinct internal models of one another. Our work extends this line by endowing each agent with an explicit Theory of Mind module that predicts the opponent’s behavior from interaction history (and, during rollouts, from simulated history), and a social EFE that directly weights the opponent's welfare, providing a step toward more cognitively grounded multi-agent active
  inference.

  Relatedly, \citet{ccatal2024belief} and \citet{pitliya2025theory} proposed a factorised active inference framework for multi-agent interactions, in which agents maintain explicit beliefs about others' internal states and preferences. Our approach is broadly congruent with this factorised perspective. Each agent maintains a separate generative model of the opponent (the ``other-model'') alongside its own self-model, and uses the other-model to simulate the opponent's expected free energy when evaluating candidate actions. \rev{The key innovation is that cooperation is organised primarily by the structural weighting of the opponent's welfare in the social EFE through the empathy parameter $\lambda$, rather than being generated by opponent learning alone. Online inference can modestly facilitate cooperation near the behavioral boundary, but does not replace this valuational contribution. The observed behavioural synchronisation and stable mutual cooperation can therefore be interpreted as coupled interaction dynamics emerging from structurally aligned social valuation.}

  Our model also connects naturally to neuroscience-inspired active inference accounts of the Theory of Mind and second-person interaction. Recent work has argued that active inference is particularly well suited to capturing the dynamics of social cognition, including mutual awareness, coordination, and reciprocal adaptation. For example, \citet{lehmann2024active} describes how second-person neuroscience can be formalised as coupled active inference processes exchanging signals over time. In our setting, such exchanges are abstracted to the observation of joint outcomes, yet these observations are sufficient to induce coupling between internal belief states through the ToM prediction mechanism. The use of two generative models with shared structure, one self-oriented and one modelling the opponent, parallels Bayesian interpretations of the mirror neuron system, according to which similar hierarchical models support both action execution and action observation.

  \rev{We can ask how much of this depends on active inference specifically, rather than on any Bayesian agent that models itself and others. The dissociation we report is not a peculiarity of the free-energy formulation. Any architecture that separates a predictive model of the partner from the weighting of the partner's outcomes in the objective should reproduce it, and related results appear in cooperative inverse reinforcement learning, where the robot infers a human reward function it is defined to share \citep{hadfield2016cooperative}, and in empowerment-based accounts, where prosocial behaviour follows from valuing the other agent's future option space \citep{salge2017empowerment}. What the active inference formulation contributes is not a different prediction but a common currency: the partner's welfare, the agent's own preferences, and the value of information about the partner are all expressed as expected free energy and traded off in a single quantity, which is what makes the epistemic term of Section~\ref{sec:methodology} and the parameter $\lambda$ commensurable. A deep Bayesian RL agent with an explicit self-and-other world model could implement the same trade-off; it would have to reintroduce the exploration bonus and the other-regarding term as separate design choices rather than deriving both from one objective.}

  \rev{Our results bear directly on the second-person picture. The phenomena that distinguish empathic dyads in our simulations, behavioural synchronisation, rapid post-defection recovery, and elevated variability near the cooperation threshold, are properties of the coupled interaction rather than of either agent's model in isolation, which is the level of description second-person neuroscience takes as primary. The model comparison in Section~\ref{sec:model_comparison} sharpens the correspondence: a ToM-only agent observes and predicts its partner but never enters a reciprocally coupled regime, whereas empathic weighting couples the two agents' free-energy landscapes and generates the interactive dynamics above. Our findings are therefore consistent with the second-person claim that engaged interaction differs from spectatorial mindreading. The correspondence has limits: our agents interact through discrete simultaneous moves rather than the continuous mutual perceptual coupling emphasised in that literature, a gap that the richer environments discussed under future directions would close.}

  \rev{This mechanism aligns with views of active inference as promoting consistency between predicted and observed social outcomes \citep{friston2018}. By incorporating the opponent's expected free energy into action evaluation, the empathic agent treats consequences for the partner as action-relevant within its own objective. When this weighting is sufficiently strong and mutual, the two agents select policies compatible with sustained joint cooperation. We describe this as \emph{empathic alignment} at the level of behavioral organization.}

  \rev{Our results also clarify the relationship between other-regarding valuation and learning-based opponent inference. The particle-filter variant progressively contracts its posterior over the opponent's behavioral profile and differentiates cooperative from predominantly defective partners. This adaptive inference modestly increases mutual cooperation when the partner is cooperative, with its largest effect near the conditional cooperation boundary. The effect remains substantially smaller than that of the focal agent's own empathic weighting, however, and weakly empathic agents remain predominantly non-cooperative despite learning that the partner is cooperative. Learning therefore refines how partner behavior is represented and can facilitate coordination, but it does not by itself supply the motivational weight assigned to the partner's welfare. The model comparison reinforces this distinction: adding opponent inversion at $\lambda=0$ produces effectively no mutual cooperation, whereas adding positive other-regarding valuation while retaining the same inversion machinery reorganizes the dyad into a cooperative regime. Because the opponent's empathic weighting is represented as a latent variable, uncertainty about the partner can also contribute an epistemic term to EFE, although isolating the behavioral contribution of that term will require a dedicated ablation.}

  \rev{The planning results reveal a more conditional relationship between strategic capability and cooperation. In symmetric-empathy mixed-horizon dyads, deeper planning produces modest reductions in mutual cooperation on the lower shoulder of the crossover, but does not shift the operational high-cooperation threshold on the tested grid. This effect reflects the weak realized reciprocity of the self-play agents: cooperating at the current round buys little additional partner cooperation in subsequent rounds, so longer rollouts slightly favor defection-initial policies. Against tit-for-tat, however, the sign reverses. Because current cooperation strongly increases the predicted probability of future partner cooperation, extending the horizon produces a large increase in both cooperation and payoff. Planning depth therefore amplifies the interaction structure encoded by the opponent model rather than exerting a uniformly anti-cooperative effect. The modest self-play reduction is reminiscent of the strategic pressure toward defection emphasized by backward-induction analyses of finitely repeated games \citep{osborne1994}, but the reversal against a reciprocating partner shows that foresight is not intrinsically equivalent to defection pressure in the present architecture.}


  \subsection{Benefits and limitations}

  \rev{In a complementary fixed-strategy stress test distinct from the planning-horizon analysis reported in Figure~\ref{fig:fig7_partner_planning}, an empathic focal agent ($\lambda_i=0.6$, opponent inversion enabled, myopic planning) interacted with each fixed strategy for $T=100$ rounds across 10 random seeds under the standard Prisoner's Dilemma payoffs. By comparison, Figure~\ref{fig:fig7_partner_planning} varies planning horizon at $\lambda_i=0.3$ over $T=60$ rounds, also across 10 seeds. In the complementary stress test, the empathic agent cooperated on 97--100\% of rounds regardless of the partner. Against unconditional defection it earns $0.00$ per round over $100$ rounds while the partner earns $5.00$, despite predicting defection almost perfectly (Brier score $0.001$). It therefore forecasts its exploiter accurately and cooperates with it anyway. Empathy as formulated here enters the agent's valuation but supplies no mechanism for withdrawing cooperation when concern is not reciprocated, leaving a high-$\lambda$ agent systematically exploitable over the tested interaction. Of the fixed strategies tested, only tit-for-tat yields a mutually beneficial outcome, with a mean payoff of $2.98$ per agent. Making the focal empathy weight conditional on the inferred $\lambda_j$, so that empathy can be regulated in response to partner disposition, is a natural next step that we have not implemented here.}

  A primary advantage of the proposed framework is its conceptual transparency and modularity. The social EFE formulation $G_{\text{social}} = (1-\lambda)G_{\text{self}} + \lambda G_{\text{other}}$ provides a single, interpretable control parameter ($\lambda$) that governs the degree of prosocial behaviour. This simplicity facilitates analysis: the cooperation threshold, exploitation dynamics, and planning-depth effects can all be understood in terms of how $\lambda$ shifts the balance between self-interest and opponent welfare in the EFE landscape. The modular architecture, separating state inference, opponent modelling (ToM and particle filter), and action selection (myopic or sophisticated), allows each component to be evaluated and improved independently.

  The opponent modelling pipeline offers additional practical benefits. The particle filter provides interpretable, online Bayesian inference over opponent profiles, and the reliability-gated blending between learned and static ToM predictions ensures graceful degradation. When insufficient data has been collected, the agent falls back to a reasonable prior rather than acting on unreliable inferences. This design pattern ``trust your model only when it has earned trust'' is broadly applicable to multi-agent systems where partner behaviour is initially unknown.

  \par \rev{The behavioral consequences of this design are substantial but conditional. When empathic weighting is sufficiently mutual, agents sustain cooperation and rapidly recover from isolated coordination failures because consequences for the partner enter their own action evaluation. The same fixed commitment creates vulnerability under asymmetric valuation, however. Other-regarding objectives can therefore support cooperative alignment, but they do not by themselves guarantee strategic robustness or protection from exploitation.}

  Nevertheless, several limitations warrant attention. First, the Theory of Mind module currently uses a static, history-conditioned payoff-based prediction of opponent responses. While the particle filter learns opponents' behavioural characteristics, the per-step opponent prediction during sophisticated planning rollouts ($t > 0$) relies on the static ToM prior, since no new observations are available during mental simulation. More sophisticated approaches, such as recursive ToM, where the opponent is modelled as itself performing ToM about the agent, could improve the fidelity of multi-step predictions, though at substantially increased computational cost.

  Second, the present implementation relies on discrete state spaces with relatively low dimensionality, and similar models. The Prisoner's Dilemma, with its four joint outcomes and two actions, is an ideal testbed for validating the core mechanism, but scaling to richer environments with continuous states, high-dimensional observations, heterogeneous models, and larger action spaces raises well-known challenges associated with policy enumeration and belief propagation. In the sophisticated planning regime, the number of candidate policies grows as $2^H$, which becomes prohibitive for large horizons. Approximate inference schemes, such as Monte Carlo tree search or amortised policy networks, may be required for more complex settings. The ultimate goal is to be able to model agents with disparate models and still enable a degree of theory of mind and empathy.

  \par \rev{The focal agent's empathic weighting $\lambda_i$ is currently fixed throughout an interaction. Although the opponent model infers the partner's latent weighting $\lambda_j$, this inference does not regulate how strongly the focal agent values the partner's welfare. Extending the framework so that the effective focal weighting $\lambda_i$ is adjusted online as a function of inferred partner disposition, interaction history, or internal state would permit strategic withdrawal under persistent exploitation and reinforcement of concern under reliable reciprocity.}

  Finally, the ethical implications of powerful social modelling must be carefully considered. The same capacities that enable an agent to cooperate effectively could also enable manipulation. An agent that accurately models another's preferences and predicts their responses could exploit this knowledge for self-serving ends when empathy is low. \rev{Our results demonstrate the underlying risk directly: low-empathy agents equipped with opponent-sensitive prediction can exploit highly empathic partners.}

\subsection{Empathy, exploitation, and the motivational gap}

The results above raise a question that the present framework can pose but not yet resolve. What separates computational empathy from genuine empathic concern?

\par \rev{The dissociation is visible most directly in the model comparison. The ToM-only agent possesses the same opponent-inversion machinery as the empathic agent but assigns no weight to the partner's welfare; symmetric ToM-only dyads remain in mutual defection, whereas empathic dyads converge to mutual cooperation. The learning analysis reaches a complementary conclusion. Updating the opponent model modestly facilitates cooperation near the behavioral boundary, but does not move weakly empathic agents into a robust cooperative regime. Social knowledge can therefore improve coordination, yet does not determine the motivational use to which that knowledge is put.}

\par \rev{The planning results further qualify this relationship. Greater lookahead is not intrinsically exploitative: it modestly reduces cooperation when the modeled partner offers little future reciprocity, but sharply increases cooperation when facing tit-for-tat. Cognitive capability magnifies the consequences represented by the agent's world and opponent models. The role of empathic weighting is to determine whether benefits and harms to the partner enter the objective being optimized. Prediction, planning, and other-regarding valuation are therefore separable components, even though their behavioral effects interact.}

\par \rev{The cognitive science literature documents analogous dissociations between social-modeling capacity and prosocial concern. Intact perspective-taking paired with weak concern for the target is often described as instrumental or ``dark'' empathy, in which social understanding supports manipulation rather than mutual benefit \citep{shamay2009two, breithaupt2019dark}. The alignment implication is not that sophisticated social modeling is inherently dangerous, but that it is motivationally underdetermined. In the present framework, the ToM module predicts the partner's behavior, whereas $\lambda$ determines whether the partner's welfare enters the focal agent's own objective.}

In the present model, $\lambda$ is fixed exogenously, a design choice that isolates the effect of empathic weighting from confounding variables. But in cognitive science, empathy is increasingly understood as a motivated capacity, dynamically regulated based on context, expected costs, and social goals rather than deployed uniformly \citep{spaulding2024motivating, zaki2014empathy}. An agent in a cooperative regime has reason to invest in allocentric modeling because it reliably reduces prediction error; an agent facing exploitation has reason to withdraw. Within active inference, this regulation maps onto precision dynamics. When allocentric predictions reliably improve model fit, their precision increases and empathic inference is upregulated; when the social environment turns adversarial, precision drops and the agent reverts toward egocentric processing. The trust-gating mechanism shown in  Section~\ref{sec:method_sophistication}, Eq.~\eqref{eq:gated_prediction} already implements a version of this logic for opponent modeling. Extending it to govern the empathy parameter itself, treating $\lambda$ as inferred rather than fixed, would allow the degree of prosocial concern to emerge from interaction dynamics. \rev{Three timescales are relevant to how such a prior could develop. Within an interaction, $\lambda$ can be regulated online from the inferred disposition of the partner; the exploitation results (Section~\ref{sec:model_comparison}: exploitability $0.98$ and a per-round payoff gap of $-4.9$ for an unconditional $\lambda=0.6$ agent facing a self-interested one) show that fixed high empathy is not stable on its own, and ongoing work in our group implements adaptive $\lambda$ regulated by the agent's own resource state. Across development, $\lambda$ can be cast as a slowly learned prior shaped by accumulated interaction outcomes, in the spirit of hierarchical active inference with fast states and slow parameters. Across generations, $\lambda$ is a trait under selection, with the payoff differences documented here supplying the selection gradient.} \rev{Two literatures constrain what such a parameter would correspond to. On the biological side, individual differences in other-regarding valuation have a documented neural substrate: computational accounts of psychopathy formulate reduced remorse and self-aggrandizement in explicitly Bayesian terms \citep{prosser2018bayesian}, and reduced structural integrity of the uncinate fasciculus tracks interpersonal psychopathy traits in men and women \citep{wolf2015interpersonal, maurer2022reduced}, with tractography in nonhuman primates suggesting the relevant limbic connectivity is evolutionarily conserved \citep{royo2025evolutionary}. Low other-regarding valuation with preserved opponent modelling offers a coarse formal analogy to this dissociation. On the developmental side, work on structure learning in self-world modelling agents describes how enduring prosocial orientations could be acquired rather than specified \citep{safron2023value}, and embodiment-based accounts argue that stable care depends on an agent having homeostatic stakes of its own \citep{christovmoore2023preventing, christovmoore2025conditions, pae2026body}. Both point the same way as our own results: because $\lambda$ is exogenous here, the model can show what other-regarding valuation does but not where it comes from.}

Such an extension, however, would not by itself ensure prosociality. Precision optimization is motivationally neutral. It determines when social modeling is useful, not whether it will be used for cooperation or exploitation. Addressing this gap likely requires agents with richer motivational architectures, systems in which prosocial behavior is grounded in something analogous to social urges (e.g., affiliation demands) whose satisfaction depends structurally on the welfare of interaction partners \cite{bach2012principles}. Integrating such motivational structure with active inference represents a natural next step toward agents whose empathy is not merely a parameter but a consequence of their own need dynamics.

\subsection{A phase-transition-like interpretation of the cooperation crossover}

\par \rev{The nonlinear cooperation boundary identified in the present model admits a restricted phase-transition-like interpretation. Empathic weighting defines a control space, while mutual cooperation provides a behavioral order parameter. Along the symmetric diagonal, $\lambda_i=\lambda_j=\lambda$, the cooperation probability rises steeply but continuously, and the analytical point $\lambda^*$ marks the maximum of the response $dP(CC)/d\lambda$. Fixed-partner sweeps provide a complementary signature: temporal variability is elevated where those paths cross the cooperation boundary and decreases as the interaction settles into consistently cooperative play.}

\par \rev{The analogy is necessarily path-dependent. The cooperation boundary is embedded in the joint parameter space $(\lambda_i,\lambda_j)$, and its apparent position depends on how that space is traversed. The conditional crossover obtained while holding the partner's empathy fixed therefore occurs at a different focal value of $\lambda_i$ from the maximal-sensitivity point on the symmetric diagonal. The system does not possess a single universal critical empathy value; rather, it contains a boundary between behavioral regimes in the dyadic control space.}

\par The binary action rule provides a simple mathematical connection to models of stochastic transitions:
\[
P_i(C)=\sigma\!\left(\beta\Delta U(\lambda)\right)
\]
Introducing the mean binary action $m=2P_i(C)-1$ gives:
\[
m=\tanh\!\left(\frac{\beta\Delta U(\lambda)}{2}\right)
\]
\rev{This has the same functional form as the mean state of a binary degree of freedom in an external field. Under the conditional-independence approximation used in the analytical treatment, $P(CC)=[(1+m)/2]^2$. As $\beta\rightarrow\infty$, the response approaches a step at $\Delta U(\lambda)=0$.}

\par \rev{The present results do not, however, establish a thermodynamic phase transition. At finite $\beta$, the logistic response remains analytic, and a two-agent system lacks both a thermodynamic limit and the collective self-consistency relation characteristic of interacting many-body models. Elevated variability near the cooperation boundary is compatible with reduced dynamical stability, but stronger claims about criticality would require targeted measurements of relaxation times and switching rates with system-size dependence.}

\par \rev{A Binder-cumulant analysis of the crossover supports this reading. Treating the per-run mean of the mutual-cooperation indicator, mapped to a $\pm1$ spin, as an order parameter $m$, the fourth-order cumulant $U_4 = 1 - \langle m^4 \rangle / (3 \langle m^2 \rangle^2)$ remains at the ordered-phase value $2/3$ on both sides of the crossover and departs from it only within a narrow window around the transition. Sweeping $\lambda$ at interaction lengths $T = 200$, $350$ and $500$ rounds (30 runs per point, $\beta = 4$, no opponent inversion, matching the analytical treatment), the dip lies at $\lambda = 0.425$, $0.425$ and $0.450$, approaching the analytical point $\lambda^{\ast} = 0.45$, while its depth diminishes with $T$ ($U_4^{\min} = 0.17$, $0.48$ and $0.55$). At a genuine second-order transition the cumulant curves for different sizes would cross near a common non-trivial value; here the departure from $2/3$ instead narrows as the interaction lengthens. This is the signature of a finite-size crossover rather than a diverging correlation scale, and it regenerates from \texttt{scripts/generate\_binder\_cumulant.py}.}

\par \rev{Extending the framework to populations of interacting agents could make this analogy testable in a stronger sense. If each agent's effective utility or beliefs depended endogenously on population-level cooperation, network structure, or a mean social field, the model could acquire the collective coupling needed to examine susceptibility scaling, metastability, and critical slowing. In the present dyadic setting, we therefore use \emph{phase-transition-like} to denote a sharp finite-precision reorganisation of behavioral stability under empathic weighting.}

  \subsection{Future directions}

  Several avenues for future work naturally follow from this framework. On the modelling side, the most immediate extension is the recursive Theory of Mind, in which the opponent is modelled as itself performing ToM about the agent. The current architecture supports depth-1 ToM (``if I do $a_i$, opponent responds with $a_j$''); extending to depth-2 (``opponent anticipates my response to their action'') would enable richer strategic reasoning. Preliminary architectural support for this exists in the codebase via the \texttt{DepthTwoToM} class, but empirical evaluation of its effects on cooperation dynamics remains for future work. In the same vein, a future research avenue entails looking into heterogeneous models, and degrees of overlap to see how we can instantiate theory of mind. 

  \rev{A second promising direction involves adaptive focal empathy. Rather than holding $\lambda_i$ fixed, the agent's effective other-regarding weighting could be regulated online as a function of its posterior over the partner's disposition, interaction history, and internal state.} An agent that observes persistent exploitation could reduce its effective empathy, implementing a principled form of empathy withdrawal that balances prosocial concern with self-protection. Conversely, an agent that observes reciprocated cooperation could increase its empathy, reinforcing cooperative dynamics. Such a mechanism would bridge the gap between empathy-based and reciprocity-based accounts of cooperation by allowing empathy itself to be shaped by experience.

  \rev{A related extension is mechanism design rather than agent design. Our dyads are anonymous and history is private, so defection carries no cost beyond the partner's response. Adding reputation, where past play is observable to third parties, changes the incentive structure directly and admits analytic treatment in the iterated Prisoner's Dilemma \citep{ciesla2025reputation}. We would expect reputation to lower the empathy threshold by making defection costly in a way that self-interested planning can itself represent, which would let us ask whether an institutional fix and an empathic fix are substitutes or complements. That question also bears on whether prosocial behaviour needs other-regarding valuation at all: deeper temporal modelling might on its own favour policies that preserve long non-zero-sum interaction, a quasi-Kantian disposition emerging from patient self-interest, as hierarchical goal architectures suggest \citep{goekoop2021higher}. Our horizon results qualify that route rather than ruling it out. In symmetric-empathy mixed-horizon dyads with weak realized reciprocity, deeper planning modestly reduces cooperation near the crossover because the rollouts compound the temptation margin without generating a compensating future response. Against tit-for-tat, however, the same planner strongly increases cooperation because present cooperation buys future cooperation. With an indefinite horizon, reputational spillovers, partner choice, or relationship continuation value, patient self-interest may therefore perform part of the work assigned to $\lambda$ here. Separating these incentive-based and genuinely other-regarding routes to cooperation is a well-posed question for richer environments.}

  \par \rev{A third direction concerns scaling beyond dyadic interaction. For $N$ interacting agents, one possible extension is:
\[
G_{\text{social},i} = (1-\lambda)G_{\text{self},i}
+ \frac{\lambda}{N-1} \sum_{j\neq i}G_{\text{other},ij}
\]
Such a model would raise questions about computational tractability, heterogeneous social weighting, in-group and out-group formation, and the stability and fairness of group-level cooperation. It would also provide the system-size dimension needed to test the collective-transition hypotheses outlined above. Scaling may ultimately require replacing separate opponent models with a shared generative architecture parameterized by agent identity or social context.}

  On the computational side, the sophisticated planner's $2^H$ policy enumeration becomes intractable for large horizons. Integrating Monte Carlo tree search, neural amortisation of policy evaluation, or hierarchical planning over temporally abstracted options could extend the effective planning horizon without exponential cost growth. Implementing the framework on neuromorphic hardware could also enable real-time empathic inference in embodied agents, such as social robots that adapt their behaviour based on ongoing inference about human states and intentions.

  Finally, empirical validation in richer environments is essential. Testing empathic agents in more complex multi-agent simulations, such as public goods games, negotiation tasks, or cooperative construction, and in human-AI interaction studies would provide critical insight into the framework's robustness and social impact. \rev{Concretely, we see a three-stage roadmap. The first stage is spatial multi-agent grid-worlds with resource acquisition and freely chosen interaction partners, which test whether $\lambda$-driven cooperation survives when defection is not a single labelled action but a pattern in a policy. The second stage is communicative tasks in which cooperation requires establishing shared reference, probing the perspective-taking component under genuine information asymmetry, which the Prisoner's Dilemma cannot. The third stage is continuous social environments with mutual perceptual coupling, where synchronisation becomes measurable in the same currency as empirical dyadic studies and which are the natural home for adaptive $\lambda$. The first two stages are implementable with the present agent architecture; the third requires continuous state spaces.} Cooperative human-AI games could assess whether empathic inference improves trust, satisfaction, and coordination relative to non-empathic baselines, providing empirical grounding for the theoretical claims advanced here.

  \section{Conclusion}\label{sec:conclusion}

\par We have presented an active inference framework in which other-regarding valuation enters action selection through the social expected free energy, $G_{\text{social},i} = (1-\lambda_i) G_{\text{self},i} + \lambda_i G_{\text{other},i}$.
Each agent combines this valuation with a history-conditioned Theory of Mind model that predicts the partner's response to candidate actions. This architecture separates two components that are often conflated under the term empathy: predicting another agent's behavior and assigning value to that agent's welfare. It thereby provides an interpretable mechanism through which cooperation can arise without explicit communication or centralized coordination, while retaining a clear distinction between social inference and prosocial motivation.

\par \rev{The Iterated Prisoner's Dilemma experiments reveal that empathic weighting reorganizes both the outcomes and temporal stability of repeated interaction. In the joint control space $(\lambda_i,\lambda_j)$, sufficiently strong and symmetric other-regarding valuation supports sustained mutual cooperation, whereas strong asymmetry exposes the more empathic agent to systematic exploitation. Along the symmetric diagonal, cooperation undergoes a sharp but continuous finite-precision crossover, with an analytical maximal-sensitivity point at $\lambda^*=0.451$ under the conditional-independence approximation. Fixed-partner sweeps show that the apparent cooperation boundary is path-dependent and that temporal variability is elevated where those paths cross it. These observations support a phase-transition-like interpretation in a restricted dynamical sense, but not the claim of a thermodynamic phase transition.}

\par \rev{Opponent inference and planning modify these dynamics without replacing their valuational basis. Online Bayesian learning contracts the posterior over partner profiles and modestly facilitates cooperation near the behavioral boundary, but weakly empathic agents remain predominantly non-cooperative even when facing a cooperative partner. Direct model comparison further shows that opponent inversion at $\lambda=0$ does not generate cooperation, whereas positive other-regarding valuation does so while retaining the same predictive machinery. Planning depth is likewise motivationally neutral: it slightly reduces cooperation when the modeled partner offers little future reciprocity, but strongly increases cooperation against a reciprocating partner such as tit-for-tat. Foresight therefore amplifies the long-run consequences represented by the opponent model rather than intrinsically favoring cooperation or defection.}

\par \rev{These results have a two-sided implication for AI alignment. Making another agent's welfare internally consequential can stabilize cooperation and recovery from coordination failures, but fixed unconditional concern also creates predictable vulnerability when partners do not reciprocate. Social prediction, planning capacity, and other-regarding valuation must therefore be designed and evaluated as distinct but interacting components. A socially capable agent is not aligned merely because it understands others, nor because it assigns them positive value under every circumstance. Robust social alignment will require mechanisms that preserve concern while regulating it in response to partner disposition, interaction history, internal needs, and broader institutional incentives.}

\par \rev{Active inference provides a useful foundation for this program because beliefs about others, preferences over outcomes, epistemic value, and prospective action can be expressed within a common objective. The present dyadic model is intentionally limited, but it establishes a tractable basis for studying how other-regarding valuation reshapes strategic interaction. Extending the framework to adaptive empathy, heterogeneous agents, richer environments, and multi-agent populations will help determine whether these local cooperation mechanisms can scale into robust forms of collective coordination. The broader objective is not simply to build agents that predict social consequences, but to build systems whose internal organization makes those consequences appropriately matter.}

\section*{Data accessibility}
\rev{All code and data supporting this article are publicly available at \url{https://github.com/mahault/empathy-prisonner-dilemma} and permanently archived on Zenodo at \url{https://doi.org/10.5281/zenodo.21908008} (release v1.0.1). The repository contains the agent implementation, the analysis scripts used for every result reported here, and the raw simulation output.

The analyses reported in this revision are reproduced by \texttt{scripts/run\_referee\_analyses.py} (model comparison, Section~\ref{sec:model_comparison}; planning horizon across payoff structures, Section~\ref{sec:sophistication}), \texttt{scripts/run\_referee2\_weights.py} (self/other weight space), and \texttt{scripts/generate\_revision\_figures.py} (Figures~\ref{fig:fig8_model_comparison}, \ref{fig:fig9_payoff_horizon} and \ref{fig:fig10_weight_space}). Raw per-run output is stored as JSON in \texttt{results/referee\_analyses/} together with a summary of the reported values. The script \texttt{scripts/verify\_referee\_analyses.py} re-runs the load-bearing results on a seed range disjoint from the one reported, and all checks pass.

Simulations are deterministic given the random seeds stated in the text, so every reported number is exactly reproducible. Reproduction requires the \texttt{si\_branch\_observations} fork of \texttt{pymdp} pinned in \texttt{environment.yml}, since the sophisticated-inference routines used here are not in the released version.}

\section*{Ethics}
\rev{This work is a simulation study. It involved no human participants, no animal subjects, and no personal data, so no ethical approval was required.}

\section*{Competing interests}
\rev{We declare we have no competing interests.}

\section*{Funding}
\rev{We received no specific funding for this work.}

\section*{Authors' contributions}
\rev{M.A. conceived and supervised the study, implemented the model, and ran the simulations and the corrected analyses. H.P. refined the results and the figures, consolidated the corrected analyses and revised the full manuscript. A.J.R. derived the analytical cooperation threshold and contributed the memory and cooperation-transition analyses. P.W., A.M., S.V.N. and H.S. contributed to the theoretical framework, the implementation and the interpretation of results. All authors critically revised the manuscript and approved the final version.}

\section*{Acknowledgements}
\rev{We thank Dalton A. R. Sakthivadivel for his contributions to the early development of this work.}

\bibliographystyle{unsrtnat}
\bibliography{main}

\end{document}